\documentclass[nofootinbib, superscriptaddress,
  preprintnumbers, amsmath, amssymb]{revtex4-2}

\usepackage[letterpaper,margin=1in]{geometry}
\usepackage{graphicx}
\usepackage{grffile}
\usepackage{slashed}
\usepackage{bbm}
\usepackage{footmisc}
\usepackage{multirow}
\usepackage{hyperref}
\hypersetup{colorlinks=true, linkcolor=blue!50!black, citecolor=green!40!black,
            urlcolor=blue!50!black}
\usepackage{amssymb}
\usepackage{amsmath}
\usepackage{color}
\usepackage{xspace}
\usepackage{caption}
\usepackage{subcaption}
\usepackage{natbib}
\usepackage{slashed}
\usepackage{xcolor}
\usepackage{placeins}

\usepackage{datetime}
\usepackage{color,fancybox}
\def\beq{\begin{equation}}
\def\eeq{\end{equation}} \def\bea{\begin{eqnarray}}
\def\eea{\end{eqnarray}}

\newcommand{\AratioFifteen}{38.7}
\newcommand{\AratioOne}{1.0}
\newcommand{\AratioSeven}{3.2}
\newcommand{\AratioTen}{9.0}
\newcommand{\AratioThirty}{578}
\newcommand{\AtbMinTanb}{7}
\newcommand{\AtbThirtyPb}{2.6}
\newcommand{\AtopOnePb}{4.00}
\newcommand{\AtopTenFb}{40.0}
\newcommand{\AtopTenTanb}{10}
\newcommand{\AtopThirtyFb}{4}
\newcommand{\BenchMass}{125.2}
\newcommand{\COneRateBandDown}{43--44}
\newcommand{\COneRateBandUp}{86--88}
\newcommand{\COneSameScaleSpread}{1.3}
\newcommand{\DistBatchSpread}{0.24}
\newcommand{\DistCombinedRuns}{13}
\newcommand{\DistEffOverFull}{1.089}
\newcommand{\DistEffPrecision}{0.014--0.015}
\newcommand{\DistEffSamples}{two samples}
\newcommand{\DistFullHoursHi}{66}
\newcommand{\DistFullHoursLo}{13}
\newcommand{\DistFullPrecision}{0.018--0.030}
\newcommand{\DistMjjMedian}{0.5--0.7}
\newcommand{\DistRunsPerPoint}{twelve}
\newcommand{\DistScaleMax}{1.9}
\newcommand{\DistSmoothMax}{0.4}
\newcommand{\DistTanbList}{1, 7, 10, 15\text{ and }30}
\newcommand{\DistXsecMinTanb}{7}
\newcommand{\DistXsecSpread}{16}
\newcommand{\Ecm}{13.6}
\newcommand{\FracGGTb}{67.3}
\newcommand{\FracGGTop}{57.3}
\newcommand{\FracGQTb}{31.3}
\newcommand{\FracGQTop}{37.9}
\newcommand{\FracGluonicTop}{51.4}
\newcommand{\FracQQTb}{1.4}
\newcommand{\FracQQTop}{4.8}
\newcommand{\GFexponent}{-5}
\newcommand{\GFmantissa}{1.1663787}
\newcommand{\HsampleRatio}{2.0}
\newcommand{\HsampleTb}{1982.2}
\newcommand{\HsampleTbErr}{0.3}
\newcommand{\HsampleTop}{1943.4}
\newcommand{\HsampleTopErr}{1.3}
\newcommand{\InclusiveTanbList}{1, 7, 10, 15\text{ and }30}
\newcommand{\InclusiveTanbMax}{30}
\newcommand{\InclusiveTanbRest}{7, 10, 15\text{ and }30}
\newcommand{\InclusiveTbRunsPerPoint}{twelve}
\newcommand{\InclusiveTopRunsPerPoint}{four}
\newcommand{\JetAlgo}{anti-$k_T$}
\newcommand{\JetPtMin}{30}
\newcommand{\JetR}{0.4}
\newcommand{\JetYMax}{4.4}
\newcommand{\MassScanEFTHighRatio}{2.9}
\newcommand{\MassScanFFHighRatio}{1.21}
\newcommand{\MassScanFFPeakRatio}{1.57}
\newcommand{\MassScanMax}{1000}
\newcommand{\MassScanN}{51}
\newcommand{\MassScanPeak}{346.6}
\newcommand{\MassScanRange}{100\text{--}1000}
\newcommand{\MassScanTanbList}{1, 7, 10, 20\text{ and }50}
\newcommand{\Mbot}{4.78}
\newcommand{\Mtop}{172.56}
\newcommand{\Mw}{80.3692}
\newcommand{\Mz}{91.1880}
\newcommand{\PDFmember}{0}
\newcommand{\PDFset}{PDF4LHC21\_mc}
\newcommand{\PhiAsymHi}{-12}
\newcommand{\PhiAsymLo}{-21}
\newcommand{\PhiAsymNullMax}{1.2}
\newcommand{\PhiMassScanBotMax}{2}
\newcommand{\PhiSumRulePull}{0.03}
\newcommand{\PhiSumRuleRatio}{1.00001}
\newcommand{\PhiSumRuleRatioErr}{0.00029}
\newcommand{\PhiTanbScanCotDev}{1}
\newcommand{\PhiTanbScanMinList}{6, 8, 15\text{ and }15}
\newcommand{\PhiTwoAmpAAAZ}{0.30}
\newcommand{\PhiTwoAmpHAAZ}{0.40}
\newcommand{\PhiTwoKeepAAZ}{46--58}
\newcommand{\RebinDetaSentence}{}
\newcommand{\RebinMjjSentence}{}
\newcommand{\RebinPhitwojjFactor}{four}
\newcommand{\RebinPtHiggsFactor}{five}
\newcommand{\RebinPtHiggsSentence}{Above 600~GeV, five native bins are combined.}
\newcommand{\RebinPtjoneSentence}{Above 400~GeV, five native bins are combined.}
\newcommand{\RebinPtjthreeSentence}{Above 110~GeV, five native bins are combined.}
\newcommand{\RebinZstarSentence}{Every two native bins are combined.}
\newcommand{\ScaleDenom}{4}
\newcommand{\ScanBandDown}{43--44}
\newcommand{\ScanBandUp}{83--89}
\newcommand{\ScanDistDevMax}{1.1}
\newcommand{\ScanPoints}{1034}
\newcommand{\ScanPrecision}{0.3--0.5}
\newcommand{\SubprocSumDevBound}{0.05}
\newcommand{\SubprocSumPullMax}{0.5}
\newcommand{\SubprocTableRunsPerPoint}{four}
\newcommand{\TailPtHiggsCut}{600}
\newcommand{\TailPtHiggsNative}{2--4}
\newcommand{\TailPtHiggsRebinned}{1--2}
\newcommand{\TailPtjthreeCut}{110}
\newcommand{\TailPtjthreeNative}{2--7}
\newcommand{\TailPtjthreeRebinned}{1--4}
\newcommand{\TanbScanCotDev}{1}
\newcommand{\TanbScanMasses}{125.2, 200, 400\text{ and }600}
\newcommand{\TanbScanMinList}{6, 8, 15\text{ and }15}
\newcommand{\TanbScanN}{20}
\newcommand{\TanbScanRange}{1\text{--}50}
\newcommand{\ThresholdRatioA}{1.94}
\newcommand{\ThresholdRatioPhi}{1.50}

\begin{document}

\title{Using Higgs Boson Production in Association with Three Jets via
  Gluon Fusion at the LHC to Investigate the $CP$ Properties of the
  Higgs Boson}

\author{Arunima Bhattacharya}
\email{arunimab@ific.uv.es}
\affiliation{Theory Division,
  IFIC, University of Valencia-CSIC, E-46100 Paterna, Valencia,
  Spain}

\author{Francisco Campanario} 
\email{francisco.campanario@ific.uv.es} 
\affiliation{Theory Division,
  IFIC, University of Valencia-CSIC, E-46100 Paterna, Valencia,
  Spain}

\author{Michael Kubocz}
\email{kubocz@physik.rwth-aachen.de}
\affiliation{III. Physics Institute B, RWTH Aachen University,
  Otto-Blumenthal-Straße 19, 52074 Aachen, Germany}

\author{Terrance Figy} \email{Terrance.Figy@wichita.edu}
\affiliation{MSP Department, Wichita State University, 1845 Fairmount
  Box 32, Wichita, KS 67206-0032, USA}

\author{Tunde Kushimo}
\email{txkushimo@shockers.wichita.edu}
\affiliation{MSP Department, Wichita State University, 1845 Fairmount
  Box 32, Wichita, KS 67206-0032, USA}

\author{Ivan Rosario}
\email{ivanro@ific.uv.es}
\affiliation{Theory Division,
  IFIC, University of Valencia-CSIC, E-46100 Paterna, Valencia,
  Spain}

\begin{abstract} 
 \noindent A neutral Higgs boson $\Phi$ with general scalar and
 pseudoscalar couplings and accompanying jets can be efficiently produced via gluon fusion
 through heavy-quark loops in high-energy hadronic collisions.  In
 this article, we present results for integrated cross sections and
 kinematic distributions for the production of a generic Higgs boson
 $\Phi$ in association with three jets, $pp \to \Phi jjj$, at order
 $\alpha_{s}^{5}$.  We include the full mass dependence of the top-
 and bottom-quark loop Feynman diagram contributions, combined
 coherently so that their interference is retained.  Earlier
 finite-mass calculations of $\Phi+3$ jets with bottom-quark loops and
 $CP$-odd or $CP$-mixed couplings were restricted to $gg \to ggg\Phi$;
 we include the complete set of
 partonic subprocesses, and find that the quark-containing channels
 supply close to one half of the pseudoscalar three-jet cross section
 at $\tan\beta=1$.  In a Type-II two-Higgs-doublet model the
 bottom-quark loop comes to dominate the rate as $\tan\beta$ is
 increased, so that the cross section is non-monotonic in $\tan\beta$.
 Comparing against the heavy-top effective theory and its
 form-factor-improved version, we show that the effective description
 fails in the hard tails of the transverse-momentum and dijet-mass
 distributions, while the third-jet centrality $z^\star$ and the
 rapidity separation of the two leading jets are largely insensitive
 to the heavy-quark treatment.  We further study the multi-jet
 azimuthal correlation $\Phi_{2}$, built from the jet momenta on the
 two sides of the Higgs boson in rapidity, and find that the
 bottom-quark contribution modifies its shape even for a state of
 definite $CP$.  For a $CP$-mixed state the interference of the two
 couplings adds a $\sin2\Phi_2$ term, which a simple counting asymmetry
 isolates and which no $CP$ eigenstate can produce.  Numerical results are presented for the LHC at
 $\sqrt{s}=\Ecm$~TeV and $m_\Phi=\BenchMass$~GeV.
\end{abstract}


\maketitle
\section{Introduction}
\label{intr}

\noindent
After the discovery of the Higgs boson by the ATLAS and CMS
collaborations at the Large Hadron Collider (LHC) in 2012
\cite{Aad:2012tfa,Chatrchyan:2012ufa}, increasingly precise
measurements have been performed to determine its properties and
interactions. Measurements of its spin, $CP$ properties and couplings
to Standard Model (SM) particles are so far consistent with the
predictions for the SM Higgs boson
\cite{ATLAS:2022vkf,CMS:2022dwd}. An important remaining question is
the $CP$ structure of the Higgs interactions. This is particularly
interesting because the amount of $CP$ violation present in the SM is
insufficient to account for the observed baryon asymmetry of the
Universe \cite{Morrissey:2012db}. Searches for additional sources of
$CP$ violation therefore form an important part of the Higgs physics
program at the LHC.

A neutral spin-zero state need not be a $CP$ eigenstate. We denote a
general state containing a $CP$-even scalar component $H$ and a
$CP$-odd pseudoscalar component $A$ by
$\Phi=H\cos\alpha+A\sin\alpha$, where $\alpha$ is the corresponding
mixing angle. Current measurements are consistent with the
predominantly $CP$-even SM expectation, while non-zero $CP$-odd
contributions to individual Higgs interactions are still allowed.
Recently, the CMS collaboration investigated anomalous $CP$-odd
contributions to the loop-induced Higgs--gluon interaction using
$H\to\gamma\gamma$ events produced through gluon fusion in association
with two jets \cite{CMS:2026yti}. The results are consistent with the
SM expectation and provide direct constraints on the $CP$ structure
of the effective Higgs--gluon interaction.

At the LHC, the SM Higgs boson is dominantly produced through
loop-induced gluon fusion. Additional hard QCD radiation gives rise
to Higgs production in association with resolved jets. Inclusive
gluon-fusion production has only limited direct sensitivity to the
tensor structure of the Higgs--gluon interaction, whereas the
kinematics of the associated jets provide characteristic angular
observables. In particular, Higgs boson production in association
with two jets through gluon fusion is well known to be sensitive to
the $CP$ properties of the Higgs--gluon interaction and, through the
heavy-quark loops, to the corresponding Higgs couplings to fermions
\cite{Plehn:2001nj,Klamke:2007cu,Campanario:2010mi}.

A characteristic observable in $\Phi+2$-jet production is the
signed azimuthal-angle difference $\phi_{jj}$ between the two tagging
jets. The characteristic distribution
for a pure $CP$-even state has minima around
$\phi_{jj}=\pm\pi/2$, whereas for a pure $CP$-odd state the minima
occur around $\phi_{jj}=0,\pm\pi$. When both scalar and
pseudoscalar couplings are present, their interference shifts the
positions of these minima according to their relative strength.
The azimuthal correlation therefore contains direct information on
the $CP$ structure of the Higgs interaction. A recent
phenomenological analysis has furthermore shown that the sensitivity
of the gluon-fusion two-jet channel can be improved by exploiting
additional kinematic information beyond the azimuthal-angle
correlation alone \cite{Bahl:2023qwk}.

An important question is whether these correlations remain robust
against additional hard QCD radiation. Matrix-element and
parton-shower studies have shown that the characteristic
$CP$-sensitive structure is preserved to a good approximation in
multi-jet events \cite{DelDuca:2006hk,Andersen:2010zx}. For such
events, Ref.~\cite{Andersen:2010zx} introduced an azimuthal
observable $\Phi_2$, constructed from the vector sums of the jet
momenta on the two sides of the Higgs boson in rapidity rather than
from the two tagging jets alone.
The angle between the
corresponding transverse momentum vectors retains the characteristic
$CP$ dependence even in the presence of additional hard radiation.
Because a third hard jet makes the two-jet variable $\phi_{jj}$ an
unnatural choice, $\Phi_2$ is the azimuthal observable we study in
$\Phi+3$-jet production below; it is defined explicitly in
Sec.~\ref{subsec:azimuthal_results}.

The radiation pattern of the third jet also provides information for
distinguishing gluon fusion from vector-boson fusion (VBF). In the
dominant VBF topology, the two scattering quark lines exchange a
color-singlet weak boson and no color is exchanged between the two
quark lines. Additional QCD radiation is therefore suppressed in the
rapidity interval between widely separated tagging jets. In
gluon-fusion production, color is instead exchanged across the
event, allowing additional radiation to populate the region between
the tagging jets. This difference has motivated studies of third-jet
rapidity distributions and central-jet vetoes in Higgs production
with several jets
\cite{Barger:1994zq,DelDuca:2004wt,DelDuca:2006hk,Campanario2013,Chen:2026vds,Chen:2021phj,Campanario:2018ppz,Figy:2007kv}.
The rapidity and transverse momentum of the third jet are consequently
important observables in Higgs-plus-three-jet production.

The calculation of Higgs production through gluon fusion becomes
increasingly demanding with the number of associated jets because the
Higgs--gluon interaction is mediated by a heavy-quark loop already at
leading order. A considerable simplification is obtained in the
large-top-mass limit, in which the top quark is integrated out and
the interaction between the Higgs boson and gluons is represented by
local effective operators. Higgs production in association with
three jets was studied in this approximation at leading order in
Ref.~\cite{DelDuca:2004wt}. NLO QCD corrections were subsequently
calculated for Higgs production with two and three jets in the same
heavy-top effective theory
\cite{Campbell:2006xx,Cullen:2013saa}. In particular, the available
fixed-order NLO QCD calculation for gluon-fusion Higgs production
with three jets employs the infinite-top-mass approximation.

Considerable progress has also been made in retaining finite
top-quark-mass effects beyond leading order at lower jet
multiplicities. For Higgs production with one jet, the NLO QCD
calculation is available with exact top-quark-mass dependence
\cite{Jones:2018hbb}. For
Higgs production with two jets, all top-quark-mass effects are
retained in the real-emission contributions, while the two-loop
virtual amplitudes are approximated by reweighting the corresponding
leading-order result with its exact mass dependence
\cite{Chen:2021azt}. Complementary high-energy resummation
approaches have also been developed for Higgs production with jets in
regions characterized by large rapidity separations
\cite{Andersen:2023kuj}. Nevertheless, for the complete
Higgs-plus-three-jet process, exact finite-quark-mass predictions are
currently restricted to leading order.

For a Higgs boson mass below the top-quark mass and sufficiently
small transverse momenta, the heavy-top approximation provides a good
description of gluon-fusion Higgs production. Its range of validity
is, however, limited. When the transverse momentum of the Higgs boson
or of one of the jets becomes comparable to or larger than the
top-quark mass, the internal structure of the loop is resolved and
the effective description begins to break down. This becomes
particularly relevant for Higgs production with several jets, where
large transverse momenta can be reached over a sizeable region of
phase space
\cite{Campanario:2013mga,Campanario:2014oua,Greiner:2016awe}.

Effective-theory predictions are often supplemented by form factors
derived from the corresponding quark-triangle amplitudes in order to
restore part of the missing quark-mass dependence. These form factors
improve the description of inclusive rates and reproduce the
characteristic heavy-quark threshold behavior. They do not,
however, contain the complete momentum dependence of the box,
pentagon and hexagon amplitudes entering multi-jet production. For
the gluonic contribution to $\Phi+3$-jet production,
Ref.~\cite{Campanario:2014oua} showed that the form-factor-improved
effective theory extends the Higgs mass range over which the total
cross section is reasonably approximated, while sizeable deviations
from the exact mass-dependent calculation remain in differential
distributions at large transverse momenta. A subsequent study of SM
Higgs production with up to three jets confirmed that the breakdown
of the heavy-top approximation is governed predominantly by the
largest transverse momentum in the event
\cite{Greiner:2016awe}.

The effective description also becomes unreliable when bottom-quark
loop contributions are enhanced. This occurs, for example, for a
pseudoscalar Higgs boson in a Type-II two-Higgs-doublet model
(2HDM). The pseudoscalar Yukawa coupling to up-type quarks is
proportional to $\cot\beta$, whereas the coupling to down-type
quarks is proportional to $\tan\beta$. Increasing $\tan\beta$
therefore suppresses the top-quark contribution and enhances the
bottom-quark contribution. At sufficiently large $\tan\beta$,
bottom-quark loops can dominate the production amplitude and
substantially modify both integrated cross sections and differential
distributions
\cite{Campanario:2010mi,Campanario:2014oua}. In this region the
exact top- and bottom-quark mass dependence of the loop amplitudes,
including their interference, is required.

Full quark-mass effects in $\Phi+2$-jet production were calculated
for the complete set of partonic subprocesses in
Ref.~\cite{Campanario:2010mi}. That calculation includes top- and
bottom-quark loops and arbitrary combinations of scalar and
pseudoscalar Yukawa couplings. For Higgs production in association
with three jets, the first finite-mass calculations concentrated on
the purely gluonic subprocess. The $CP$-even contribution
$gg\to gggH$ was studied in Ref.~\cite{Campanario:2013mga}, and
the calculation was subsequently extended in
Ref.~\cite{Campanario:2014oua} to a general neutral Higgs state with
scalar and pseudoscalar couplings. The latter calculation included
both top- and bottom-quark loops and their interference, but its
phenomenological analysis was restricted to the subprocess
$gg\to ggg\Phi$.

The purely gluonic subprocess contains massive rank-five hexagon
integrals, which are the highest-point and highest-rank loop
topologies entering Higgs-plus-three-jet production at this order.
The associated numerical instabilities were controlled using
Ward-identity tests and a rescue system based on quadruple precision
\cite{Campanario:2013mga,Campanario:2014oua}. This procedure was
shown to remain stable also for configurations dominated by
bottom-quark loops. For a SM $CP$-even Higgs boson, the complete set
of partonic channels for Higgs production with up to three jets was
subsequently calculated at leading order while retaining the full
top- and bottom-quark mass dependence
\cite{Greiner:2016awe}.

The loop-induced $\Phi+3$-jet process is also included in
VBFNLO~3.0 at leading order for $CP$-even and $CP$-odd scalar states,
retaining the full top-quark-mass dependence
\cite{Baglio:2024gyp}. In the present work, we perform a detailed
study of the complete $\Phi+3$-jet process within the GGFLO
implementation of VBFNLO and extend the $CP$-sensitive and
bottom-quark analysis of Ref.~\cite{Campanario:2014oua} from the
purely gluonic subprocess to the complete set of partonic channels.

We calculate $pp\to\Phi jjj$
at $\mathcal{O}(\alpha_s^5)$. All gluon-, quark-gluon- and
quark-initiated subprocesses, together with their crossing-related
channels, are included. The exact top- and bottom-quark mass
dependence is retained in the one-loop amplitudes, and the two
heavy-quark contributions are combined coherently. The formalism
allows independent scalar and pseudoscalar Yukawa couplings and
therefore covers pure $CP$-even, pure $CP$-odd and genuinely
$CP$-mixed configurations. We refer to a state as mixed only when
both its scalar and pseudoscalar amplitudes are non-zero.

We investigate integrated cross sections for the individual partonic
subprocesses and for their sum, compare top-only predictions with
predictions containing both top- and bottom-quark loops, and study the
dependence of the pseudoscalar sector on $\tan\beta$. We also study
differential distributions in the transverse momenta of the Higgs
boson and the jets, the dijet invariant mass and rapidity separation,
and observables describing the position of the third jet. Particular
attention is given to azimuthal correlations relevant for $CP$
measurements and to their behavior in the presence of the additional
hard jet. Numerical results are presented for a Higgs boson mass of
$\BenchMass$~GeV at the LHC center-of-mass energy
$\sqrt{s}=\Ecm$~TeV.

The remainder of the paper is organized as follows. In
Section~\ref{sec:theo}, we introduce the theoretical framework and
the effective-theory approximation. In Section~\ref{sec:calc}, we
describe the construction of the one-loop amplitudes, their
implementation in VBFNLO and the numerical checks. The integrated
cross sections and differential distributions are presented in
Section~\ref{sec:results}. We summarize our conclusions in
Section~\ref{sec:summ}.

\section{Theoretical Framework}
\label{sec:theo}

\noindent
We consider a neutral Higgs sector containing a $CP$-even scalar
$H$ and a $CP$-odd pseudoscalar $A$. Their Yukawa interactions with
quarks are written as
\begin{equation}
\label{eq:yukawa}
\mathcal{L}_{\rm Yuk}
=
\sum_q
\left(
y_q\,\bar q q\,H
+
i\,\widetilde y_q\,\bar q\gamma_5 q\,A
\right),
\end{equation}
where $y_q$ and $\widetilde y_q$ denote the scalar and pseudoscalar
Yukawa couplings, respectively. In the loop-induced amplitudes
considered below, the massive quarks circulating in the Higgs
production loop are the third-generation quarks, $q=t,b$.

In the presence of $CP$ mixing, the $CP$ eigenstates $H$ and $A$
may be related to the mass eigenstates $\Phi$ and $\Phi'$ by
\begin{equation}
\label{eq:mixing}
\begin{pmatrix}
\Phi \\[1mm]
\Phi'
\end{pmatrix}
=
\begin{pmatrix}
\cos\alpha & \sin\alpha \\
-\sin\alpha & \cos\alpha
\end{pmatrix}
\begin{pmatrix}
H \\[1mm]
A
\end{pmatrix},
\end{equation}
where $\alpha$ denotes the mixing angle. We assume that the
orthogonal state $\Phi'$ is sufficiently heavy that its contribution
to the processes considered here can be neglected. Inverting
Eq.~\eqref{eq:mixing} and retaining only the light state $\Phi$, the
Yukawa interaction becomes
\begin{equation}
\label{eq:yukawa_phi}
\mathcal{L}_{\rm Yuk}
=
\sum_q
\bar q
\left(
Y_q+i\gamma_5\widetilde Y_q
\right)
q\,\Phi,
\end{equation}
with
\begin{equation}
\label{eq:mixed_yukawa}
Y_q=y_q\cos\alpha,
\qquad
\widetilde Y_q=\widetilde y_q\sin\alpha.
\end{equation}
The limits $\alpha=0$ and $\alpha=\pi/2$ correspond to pure
$CP$-even and pure $CP$-odd states, respectively
\cite{Campanario:2010mi,Campanario:2014oua}.

At amplitude level, the contribution of a general state can therefore
be written schematically as
\begin{equation}
\label{eq:mixed_amplitude}
\mathcal{M}_{\Phi}
=
\sum_{f=t,b}
\left(
Y_f\,\mathcal{M}^{H}_f
+
\widetilde Y_f\,\mathcal{M}^{A}_f
\right),
\end{equation}
where $\mathcal{M}^{H}_f$ and $\mathcal{M}^{A}_f$ denote the scalar
and pseudoscalar loop amplitudes for a quark of mass $m_f$.
A genuinely $CP$-mixed configuration requires at least one non-zero
scalar coupling $Y_f$ and at least one non-zero pseudoscalar coupling
$\widetilde Y_f$. An overall rescaling of a purely scalar or purely
pseudoscalar amplitude does not by itself constitute $CP$ mixing.

In the SM, the gluon-fusion amplitude is dominated by the top-quark
loop, while the bottom-quark contribution is suppressed both by its small
Yukawa coupling, $y_b^{\rm SM}=m_b/v$, with $v\simeq246$~GeV, and by an
additional factor of $m_b$ from the mass insertion required to compensate
the chirality flip at the Higgs--quark vertex. Its amplitude relative to the top loop
scales as $(m_b^2/m_H^2)\ln^2(m_H^2/m_b^2)$. In the limit in which the top-quark mass is much
larger than all kinematic scales entering the process, the top quark
can be integrated out. The interactions of the scalar and
pseudoscalar fields with gluons can then be written as
\begin{equation}
\label{eq:Leff_HA}
\mathcal{L}_{\rm eff}
=
y_t\,\frac{\alpha_s}{12\pi m_t}\,
H\,G_{\mu\nu}^{a}G^{a\,\mu\nu}
+
\widetilde y_t\,\frac{\alpha_s}{8\pi m_t}\,
A\,G_{\mu\nu}^{a}\widetilde G^{a\,\mu\nu},
\end{equation}
where $G_{\mu\nu}^{a}$ is the gluon field-strength tensor and
\begin{equation}
\label{eq:dual_field_strength}
\widetilde G^{a\,\mu\nu}
=
\frac{1}{2}\,
\epsilon^{\mu\nu\rho\sigma}
G_{\rho\sigma}^{a},
\qquad
\epsilon^{0123}=+1.
\end{equation}
For the mixed state of Eq.~\eqref{eq:mixing}, the corresponding
effective interaction is
\begin{equation}
\label{eq:Leff_phi}
\mathcal{L}_{\rm eff}
=
\left[
Y_t\,\frac{\alpha_s}{12\pi m_t}\,
G_{\mu\nu}^{a}G^{a\,\mu\nu}
+
\widetilde Y_t\,\frac{\alpha_s}{8\pi m_t}\,
G_{\mu\nu}^{a}\widetilde G^{a\,\mu\nu}
\right]\Phi.
\end{equation}
For Yukawa couplings proportional to the quark mass, the explicit
factor of $m_t$ in the denominator cancels against the mass contained
in the Yukawa coupling. The resolved heavy-quark loop is thereby
replaced by a local Higgs--gluon interaction
\cite{Spira1998a,Campanario:2010mi,Campanario:2014oua}.

The relative normalization of the two operators in
Eq.~\eqref{eq:Leff_HA} is important for $CP$-mixed configurations.
For equal scalar and pseudoscalar Yukawa couplings, the coefficient of
the pseudoscalar operator is larger by a factor of $3/2$. Thus equal
numerical values of $Y_t$ and $\widetilde Y_t$ do not correspond to
equal scalar and pseudoscalar effective gluonic amplitudes in the
heavy-top limit. For example, the phenomenological benchmark used in
Ref.~\cite{Campanario:2014oua} was defined by
\begin{equation}
\label{eq:old_mixed_benchmark}
y_q=\widetilde y_q=\widetilde y_q^{\,\rm II},
\qquad
\tan\alpha=\frac{2}{3},
\end{equation}
so that the factor $3/2$ in the pseudoscalar effective coupling is
compensated and the scalar and pseudoscalar components have equal
strength in the heavy-top limit. Equation~\eqref{eq:old_mixed_benchmark}
is a phenomenological benchmark and should not be interpreted as the
generic scalar-coupling structure of a Type-II 2HDM.

Finite-quark-mass effects can be incorporated approximately into the
effective-theory prediction using form factors derived from the
corresponding one-loop triangle amplitudes. For a scalar or
pseudoscalar Higgs coupling to a quark of mass $m_i$, the form factors
are
\begin{align}
\label{eq:form_factors}
{\rm FF}_{A}(\tau_i)
&=
2\,\tau_i^{-1}f(\tau_i),
\\
{\rm FF}_{H}(\tau_i)
&=
2\left[
\tau_i+(\tau_i-1)f(\tau_i)
\right]\tau_i^{-2},
\end{align}
where
\begin{equation}
\label{eq:f_tau}
f(\tau)
=
\begin{cases}
\displaystyle
\arcsin^2\!\sqrt{\tau},
& \tau\leq1,
\\[3mm]
\displaystyle
-\frac{1}{4}
\left[
\ln\!\left(
\frac{1+\sqrt{1-\tau^{-1}}}
     {1-\sqrt{1-\tau^{-1}}}
\right)
-i\pi
\right]^2,
& \tau>1,
\end{cases}
\end{equation}
and
\begin{equation}
\label{eq:tau}
\tau_i=\frac{m_\Phi^2}{4m_i^2},
\qquad
i=t,b.
\end{equation}
The point $\tau_i=1$ corresponds to the threshold
$m_\Phi=2m_i$. In the EFT+FF predictions of this work, the effective
coupling of Eq.~\eqref{eq:Leff_HA} is multiplied by the top-quark form
factor alone, normalized to its heavy-quark limit,
${\rm FF}(\tau_t)/{\rm FF}(0)$ with ${\rm FF}_A(0)=2$ and ${\rm FF}_H(0)=4/3$,
evaluated at $\tau_t=m_\Phi^2/(4m_t^2)$; no bottom-quark contribution enters
EFT+FF. These form factors restore the quark-mass dependence
of the corresponding triangle amplitude and reproduce its threshold
structure. They do not, however, contain the complete momentum
dependence of the box, pentagon and hexagon contributions entering
$\Phi+3$-jet production. Consequently, the form-factor-improved
effective theory is used below only as an approximation with which
the exact mass-dependent calculation can be compared
\cite{Spira1998a,Campanario:2014oua}.

The importance of retaining the bottom-quark mass dependence becomes
particularly clear in extended Higgs sectors. For the pseudoscalar
component of a Type-II 2HDM, the Yukawa couplings to up- and
down-type quarks are
\begin{equation}
\label{eq:typeII_yukawa}
\widetilde y^{\,\rm II}_{u}
=
-\frac{m_u}{v}\cot\beta,
\qquad
\widetilde y^{\,\rm II}_{d}
=
-\frac{m_d}{v}\tan\beta,
\end{equation}
where
\begin{equation}
\label{eq:vevs}
v=\sqrt{v_u^2+v_d^2},
\qquad
\tan\beta=\frac{v_u}{v_d}.
\end{equation}
Thus, increasing $\tan\beta$ suppresses the pseudoscalar coupling to
up-type quarks while enhancing the coupling to down-type quarks.
Bottom-quark loops can consequently become sizeable and eventually
dominate the pseudoscalar production amplitude at large
$\tan\beta$. In this region the heavy-top approximation does not
provide a reliable description and the exact top- and bottom-quark
loop amplitudes must be retained
\cite{Campanario:2010mi,Campanario:2014oua}. We emphasize that
Eq.~\eqref{eq:typeII_yukawa} specifies the pseudoscalar Yukawa
couplings. The scalar couplings $y_q$ in a general 2HDM depend on the
parameters of the scalar sector and must be specified independently.

\subsection{The mixed state evaluated in this work}
\label{subsec:mixed_state}

Equations~\eqref{eq:yukawa_phi}--\eqref{eq:mixed_yukawa} parametrize a
general mixed state. The results of Sec.~\ref{subsec:phi_distributions}
continue the benchmark of Eq.~\eqref{eq:old_mixed_benchmark}: the scalar and
pseudoscalar Yukawa couplings are given the \emph{same} Type-II structure,
\begin{equation}
\label{eq:phi_superposition}
\Phi=c_H\,H+c_A\,A ,
\qquad
y_q=\widetilde y_q=\widetilde y^{\,\rm II}_q ,
\end{equation}
so that, normalized to the Standard Model Yukawa coupling
$y_q^{\rm SM}=m_q/v$, and up to an overall sign common to all four, which
cancels in every observable, the four couplings entering
Eq.~\eqref{eq:mixed_amplitude} are
\begin{equation}
\label{eq:phi_couplings}
\begin{aligned}
\frac{Y_t}{y_t^{\rm SM}}&=c_H\cot\beta,
&\qquad
\frac{Y_b}{y_b^{\rm SM}}&=c_H\tan\beta,
\\[1mm]
\frac{\widetilde Y_t}{y_t^{\rm SM}}&=c_A\cot\beta,
&\qquad
\frac{\widetilde Y_b}{y_b^{\rm SM}}&=c_A\tan\beta .
\end{aligned}
\end{equation}
The ratio $\widetilde Y_q/Y_q=c_A/c_H$ is the same for both quarks and
independent of $\tan\beta$, so the $CP$ composition of the Yukawa couplings
is fixed. The relative strength and phase of the induced scalar and
pseudoscalar gluonic amplitudes can nevertheless depend on the quark mass
and kinematics. The choice $c_A=2/3$ compensates the relative factor $3/2$
between the operators in Eq.~\eqref{eq:Leff_HA} in the heavy-top limit.
For the on-shell bottom-quark triangle,
$|{\rm FF}_A/{\rm FF}_H|\simeq1.05$ at $m_\Phi=\BenchMass$~GeV. This
illustrates a change in relative strength, but the triangle form factors
do not define a universal mixing angle for the full $\Phi+3$-jet amplitude.
The mixing angle $\alpha$ does not appear in
Eq.~\eqref{eq:phi_couplings}: Eq.~\eqref{eq:phi_couplings} assigns the Type-II
\emph{pseudoscalar} structure to the $CP$-even component as well, which is a
deliberate benchmark choice and not the scalar sector of a Type-II 2HDM, as
already emphasized below Eq.~\eqref{eq:old_mixed_benchmark}.

We evaluate Eq.~\eqref{eq:phi_couplings} with
\begin{equation}
\label{eq:phi_benchmark}
c_H=1,
\qquad
c_A=\frac{2}{3} .
\end{equation}
The value $c_A=2/3$ plays the role of $\tan\alpha=2/3$ in
Eq.~\eqref{eq:old_mixed_benchmark}. Note that $c_H$ and $c_A$ multiply the two amplitudes directly rather than
being the components $\cos\alpha$ and $\sin\alpha$ of a normalized mixing;
relative to the unit-normalized form of Eq.~\eqref{eq:old_mixed_benchmark}
the state of Eq.~\eqref{eq:phi_benchmark} therefore carries an overall factor
$1/\cos\alpha=\sqrt{13}/3$ in the amplitude, and $13/9$ in the rate. This
affects the overall normalization of the $\Phi$ cross sections quoted below
and cancels in every normalized distribution.

Both components of the produced state are evaluated at its mass $m_\Phi$:
in the calculation all neutral-Higgs mass parameters are set to $m_\Phi$.
This is a technical input for the state that is produced, not a statement
about the spectrum; the orthogonal state $\Phi'$ of Eq.~\eqref{eq:mixing}
remains heavy and is not produced. Equations~\eqref{eq:phi_superposition}
and \eqref{eq:phi_benchmark} therefore define a phenomenological benchmark
chosen to expose the $CP$ structure, not a point of a specific
two-Higgs-doublet model, and no constraint from the scalar potential or
from the observed $125$~GeV state is imposed on it.

Our main predictions are evaluated using the exact
mass-dependent one-loop amplitudes. We distinguish predictions
containing top-quark loops only from those in which the top- and
bottom-quark amplitudes are added coherently, including their
interference. The heavy-top effective theory and its
form-factor-improved version are used for comparison where indicated.

\section{Calculation Details}
\label{sec:calc}

\noindent
We consider the production of a neutral Higgs state $\Phi$ in
association with three jets through the loop-induced gluon-fusion
mechanism at leading order,
$\mathcal{O}(\alpha_s^5)$. A convenient set of basic partonic
subprocesses is
\begin{equation}
\label{eq:3jsubproc}
\begin{array}{rclcrcl}
q\bar q &\to& q\bar q g\,\Phi,
&&
q\bar Q &\to& q\bar Q g\,\Phi,
\\
qg &\to& qgg\,\Phi,
&&
gg &\to& ggg\,\Phi,
\end{array}
\end{equation}
together with all crossing-related processes. Here $q$ and $Q$
denote quark flavors, with $q\neq Q$ in the second subprocess.
Equivalent identical- and non-identical-flavor configurations and
all corresponding crossed initial states are included in the hadronic
calculation. The Higgs boson couples to the external partons through
a closed heavy-quark loop, for which the top- and bottom-quark
contributions are retained.

The purely gluonic subprocess $gg\to ggg\Phi$, which contains the
highest-point loop topologies of the calculation, was discussed in
detail in
Refs.~\cite{Campanario:2013mga,Campanario:2014oua}. The complete
calculation extends this construction to the quark-containing
subprocesses in Eq.~\eqref{eq:3jsubproc} and to all of their
crossing-related channels.

\begin{figure}[!htbp]
  \centering
  \includegraphics[width=0.55\textwidth]{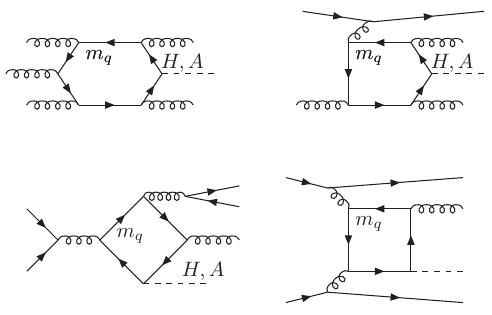}
  \caption{Representative Feynman diagrams contributing to
  $\Phi+3$-jet production with zero, two and four external quarks. The Higgs boson couples to the external partons through a
  closed heavy-quark loop.}
  \label{fig:diag_sub}
\end{figure}

Representative diagrams for the different partonic channels are
shown in Fig.~\ref{fig:diag_sub}. The maximum loop multiplicity
depends on the number of external quarks. The purely gluonic
subprocess contains topologies up to hexagons, subprocesses with two
external quarks contain topologies up to pentagons, and those
with four external quarks contain topologies up to boxes.
Lower-point loop topologies also contribute to the corresponding
partonic classes. Additional $s$-channel configurations arise in
appropriate crossing-related subprocesses, for example
$q\bar q\to ggg\Phi$.

The amplitudes are constructed using the helicity formalism and the
effective-current approach
\cite{Hagiwara:1988pp,Campanario:2011cs}. In this formulation,
external quark lines and gluons attached to the heavy-quark loop are
represented by effective currents. The complete set of Feynman
diagrams can then be assembled from a reduced set of master loop
objects. For each $CP$ assignment there are four master topologies:
a two-current vertex, a box, a pentagon and a hexagon. We denote them
by
\begin{equation}
\label{eq:master_topologies}
{\cal V}^{\varphi}_{\mu_1\mu_2},
\qquad
{\cal B}^{\varphi}_{\mu_1\mu_2\mu_3},
\qquad
{\cal P}^{\varphi}_{\mu_1\ldots\mu_4},
\qquad
{\cal H}^{\varphi}_{\mu_1\ldots\mu_5},
\qquad
\varphi=H,A.
\end{equation}
The scalar and pseudoscalar contributions therefore require eight
master loop objects in total.

\begin{figure}[!htbp]
  \centering
  \includegraphics[width=0.55\textwidth]{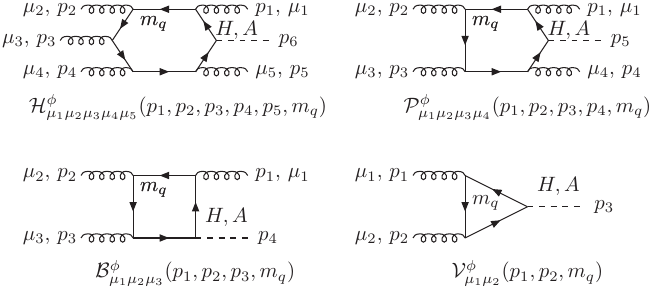}
  \caption{Master one-loop topologies used to construct the
  $\Phi+3$-jet amplitudes. For each topology, the Higgs insertion can
  correspond to a $CP$-even scalar $H$ or a $CP$-odd pseudoscalar
  $A$.}
  \label{fig:diagrams}
\end{figure}

As an illustration, consider a contribution containing one external
gluon and two effective quark currents, corresponding to a box
topology. Suppressing overall coupling and color factors, its
amplitude can be written schematically as
\begin{equation}
\label{eq:box_amplitude}
{\cal M}^{\Phi}_{\rm box}
\propto
\epsilon^{\mu_1}(p_1)\,
J_Q^{\mu_2}\,
J_q^{\mu_3}
\sum_{f=t,b}
\left[
Y_f\,
{\cal B}^{H}_{\mu_1\mu_2\mu_3}
(p_1,q_2,q_3;m_f)
+
\widetilde Y_f\,
{\cal B}^{A}_{\mu_1\mu_2\mu_3}
(p_1,q_2,q_3;m_f)
\right].
\end{equation}
Here $J_q^\mu$ denotes an effective quark current
\cite{DelDuca:2001fn}, while $Y_f$ and $\widetilde Y_f$ are the
scalar and pseudoscalar Yukawa couplings introduced in
Section~\ref{sec:theo}. The scalar and pseudoscalar master functions
are defined using
\begin{equation}
\label{eq:Gamma_HA}
\Gamma_H=1,
\qquad
\Gamma_A=i\gamma_5.
\end{equation}

For one orientation of the fermion loop, a representative definition
of the box master integral, with the Yukawa coupling and overall
normalization factors stripped off, is
\begin{equation}
\label{eq:tensor_box}
{\cal B}^{\varphi}_{\mu_1\mu_2\mu_3}
(p_1,p_2,p_3;m_f)
=
\int\frac{d^4\ell}{(2\pi)^4}\,
\frac{
N^{\varphi}_{\mu_1\mu_2\mu_3}(\ell)
}{
D_0D_1D_2D_3
},
\end{equation}
where
\begin{equation}
\label{eq:box_numerator}
\begin{aligned}
N^{\varphi}_{\mu_1\mu_2\mu_3}(\ell)
={\rm Tr}\Big[
&(\slashed{\ell}+m_f)\,
\Gamma_\varphi\,
(\slashed{\ell}+\slashed{p}_{123}+m_f)\,
\gamma_{\mu_3}
\\
&\times
(\slashed{\ell}+\slashed{p}_{12}+m_f)\,
\gamma_{\mu_2}
(\slashed{\ell}+\slashed{p}_{1}+m_f)\,
\gamma_{\mu_1}
\Big],
\end{aligned}
\end{equation}
and
\begin{equation}
\label{eq:box_denominators}
\begin{aligned}
D_0&=\ell^2-m_f^2+i0,
&
D_1&=(\ell+p_1)^2-m_f^2+i0,
\\
D_2&=(\ell+p_{12})^2-m_f^2+i0,
&
D_3&=(\ell+p_{123})^2-m_f^2+i0,
\end{aligned}
\end{equation}
with
\begin{equation}
\label{eq:p12p123}
p_{12}=p_1+p_2,
\qquad
p_{123}=p_1+p_2+p_3.
\end{equation}
For the master integral in Eqs.~(23)--(26), all momenta attached to the
loop are defined as incoming. This is a local momentum convention;
physical scattering channels are obtained by crossing. Momentum
conservation therefore gives
\begin{equation}
p_\Phi=-(p_1+p_2+p_3).
\end{equation}
The contribution with the opposite orientation of the fermion loop
is combined with Eq.~\eqref{eq:tensor_box} according to charge
conjugation.

The one-loop amplitudes are ultraviolet finite and contain no
infrared poles associated with the massive internal top- and
bottom-quark propagators. No ultraviolet renormalization of these
one-loop amplitudes is therefore required. The partonic
$\Phi+3$-parton cross section nevertheless contains the usual soft
and collinear regions associated with massless external QCD partons.
These regions are rendered finite at leading order by the
infrared-safe jet definition and the finite jet-selection cuts
specified in Section~\ref{sec:results}. Following
Refs.~\cite{Campanario:2010mi,Campanario:2014oua}, the
pseudoscalar loop amplitudes are evaluated in four dimensions using
an anticommuting $\gamma_5$ prescription.

The tensor integrals are evaluated following the implementation
described in Ref.~\cite{Campanario:2011cs}. Tensor integrals up to
four-point functions are reduced using the Passarino-Veltman
decomposition \cite{Passarino:1978jh}. Five- and six-point tensor
integrals are evaluated using the Denner-Dittmaier reduction
\cite{Denner:2005nn}, together with the implementation of
Ref.~\cite{Campanario:2011cs}. Scalar one-loop integrals are
evaluated following
Refs.~\cite{tHooft:1978jhc,Denner:1991qq}. Charge-conjugation-related
fermion-loop diagrams are combined using Furry's theorem, reducing
the number of loop diagrams that have to be evaluated. The color
algebra is implemented analytically and has been cross-checked
against MadGraph.

A central issue in the numerical evaluation is the stability of the
high-rank pentagon and hexagon tensor integrals. We monitor their
numerical stability using Ward identities. Writing the fermion
propagator as
\begin{equation}
S_f(k)=\frac{1}{\slashed{k}-m_f},
\end{equation}
the elementary identity
\begin{equation}
\label{eq:ward_propagator}
S_f(k)\,\slashed{p}\,S_f(k+p)
=
S_f(k)-S_f(k+p)
\end{equation}
relates an $N$-point loop function to a difference of
$(N-1)$-point functions when a gluon current is replaced by its
momentum. For example, contraction of a rank-five hexagon master
integral gives schematically
\begin{equation}
\label{eq:ward_hexagon}
p_i^{\mu_i}
{\cal H}^{\varphi}_{\mu_1\ldots\mu_i\ldots\mu_5}
=
{\cal P}^{\varphi,(i)}_
{1,\mu_1\ldots\widehat{\mu_i}\ldots\mu_5}
-
{\cal P}^{\varphi,(i)}_
{2,\mu_1\ldots\widehat{\mu_i}\ldots\mu_5},
\qquad
i=1,\ldots,5,
\end{equation}
where $\widehat{\mu_i}$ denotes the Lorentz index removed by the
contraction. Analogous identities are constructed for the lower-point
master integrals.

All available Ward identities are evaluated for every physical
permutation of each diagram and for every phase-space point in the
stability test. Following
Refs.~\cite{Campanario:2013mga,Campanario:2014oua}, we require a
relative accuracy of
\begin{equation}
\label{eq:gauge_tolerance}
\epsilon=5\times10^{-4}.
\end{equation}
If this criterion is not fulfilled in double precision, the scalar
integrals and tensor coefficients of the affected diagram are
re-evaluated in quadruple precision. If the Ward identities are
still not satisfied, the complete amplitude at that phase-space point
is assigned zero weight. In the previous gluonic calculation, the
fraction of points rejected after the quadruple-precision rescue was
well below the per-mille level, including configurations dominated by
bottom-quark loops
\cite{Campanario:2013mga,Campanario:2014oua}.

The complete calculation is implemented in the GGFLO component of a
development version of VBFNLO~3.0. In VBFNLO the
Higgs-plus-three-jet gluon-fusion process is identified as
process~4200. The public VBFNLO~3.0 release contains the
one-loop-induced $\Phi+3$-jet process at leading order for
$CP$-even and $CP$-odd scalar states with full top-quark-mass
dependence, and provides an OpenMPI-parallelized implementation
\cite{Baglio:2024gyp}. The implementation used in the present
work additionally retains the bottom-quark loop contribution and its
interference with the top-quark contribution in all partonic
subprocesses considered here.

Several independent checks have been performed on the complete
implementation. The heavy-top-limit amplitudes were implemented for
each partonic subprocess and compared with MadGraph for both
$CP$-even and $CP$-odd Higgs states. Agreement was obtained at
machine precision for individual amplitudes and at the per-mille
level for integrated cross sections. The exact one-loop amplitudes
were subsequently evaluated with
\begin{equation}
m_t=5\times10^4~{\rm GeV}
\end{equation}
and compared with the corresponding heavy-top-limit amplitudes,
yielding agreement at the per-mille level or better for the complete
set of subprocesses. For the purely gluonic subprocess, the more
precise comparison reported in
Ref.~\cite{Campanario:2014oua} gives agreement better than
$10^{-4}$.

As an internal consistency check, setting equal top- and bottom-quark
masses and Yukawa couplings reproduces four times the single-quark
contribution to the squared amplitude, for all partonic subprocess
classes and for both scalar and pseudoscalar couplings.

The results presented here were obtained with a development version of
VBFNLO in which two bugs have been corrected: one in the four-quark
subprocesses, and one in the user-defined scalar--pseudoscalar mixing
mode for process~4200. These corrections are not included in the public
VBFNLO~3.0 release~\cite{Baglio:2024gyp} and will be incorporated into a
forthcoming release.

\section{Numerical Results}
\label{sec:results}

\subsection{Numerical setup}
\label{subsec:numsetup}

We present numerical predictions for the loop-induced production of a
neutral Higgs state in association with three jets at the LHC. The
results in this section are obtained at a center-of-mass energy
$\sqrt{s}=\Ecm$~TeV. Unless stated otherwise, the Higgs boson mass is
fixed to $m_{\Phi}=\BenchMass$~GeV.

Jets are reconstructed with the \JetAlgo{} algorithm using a radius
parameter $R=\JetR$. They are required to satisfy
$$
p_{T,j}>\JetPtMin~{\rm GeV},
\qquad
|y_j|<\JetYMax.
$$
The jets are ordered according to transverse momentum,
$p_{T,j_1}>p_{T,j_2}>p_{T,j_3}$. Since all external light quarks and
gluons are treated as massless, jet rapidity and pseudorapidity
coincide at parton level.

The factorization and renormalization scales are set equal and chosen
dynamically as
$$
\mu_F=\mu_R=\frac{H_T}{\ScaleDenom},
$$
with
$$
H_T=E_{T,\Phi}+\sum_j p_{T,j},
\qquad
E_{T,\Phi}=\sqrt{m_\Phi^2+p_{T,\Phi}^2}.
$$
We use the \PDFset{} parton distributions through LHAPDF6
\cite{PDF4LHCWorkingGroup:2022cjn,Buckley:2014ana}. The strong
coupling is taken consistently from the PDF set. The electroweak
parameters are defined in the $G_F$ scheme, with
$$
G_F=\GFmantissa\times10^{\GFexponent}~{\rm GeV}^{-2},
\qquad
M_W=\Mw~{\rm GeV},
\qquad
M_Z=\Mz~{\rm GeV}.
$$
The heavy-quark masses entering the loop amplitudes are
$$
m_t=\Mtop~{\rm GeV},
\qquad
m_b=\Mbot~{\rm GeV}.
$$
Both pole masses are used in the loop propagators and Yukawa couplings.
Unlike Ref.~\cite{Campanario:2014oua}, the bottom Yukawa is not evolved to
the hard scale. The corresponding mass-scheme dependence is not
included in the quoted uncertainties. All other quarks are treated as
massless.

For the exact calculation we distinguish two loop configurations.
The ``top'' result contains only the top-quark amplitude, while the
``$t+b$'' result contains the coherent sum
$$
{\cal M}_{t+b}={\cal M}_t+{\cal M}_b
$$
and therefore includes top--bottom interference. The uncertainties
quoted for the integrated cross sections are Monte Carlo integration
errors only.

The scale dependence is estimated from a seven-point variation of
$(\mu_R,\mu_F)$ around the central choice, by factors of two with
$1/2\le\mu_R/\mu_F\le2$, computed in the same runs by reweighting each phase-space
point. With five powers of $\alpha_s$ at leading order it is large: the rates of
the differential samples move up by \COneRateBandUp\% and down by
\COneRateBandDown\%, and those of the $\ScanPoints$ points of the parameter
scans by \ScanBandUp\% and \ScanBandDown\%. Because the variation is close to
a common factor, it largely cancels in normalized distributions and in ratios
taken at a common scale; Sec.~\ref{subsec:finite_mass_distributions} quantifies
this. PDF uncertainties are not included. The runs of this section were set
up, run and analyzed, and the scale variation implemented, with the
assistance of an AI coding agent under the authors' direction, as detailed in
the Acknowledgments.

\subsection{Inclusive cross sections and bottom-quark effects}
\label{subsec:inclusive_results}

We first consider the pseudoscalar state $A$ at
$m_A=\BenchMass$~GeV. Table~\ref{tab:A-inclusive} gives the inclusive
cross section for the top-only and coherent top-plus-bottom
calculations as a function of $\tan\beta$.

\begin{table}[t]
\centering
\caption{
Inclusive LO cross sections for $pp\to A+3j$ at
$\sqrt{s}=\Ecm$~TeV and $m_A=\BenchMass$~GeV, at the $\tan\beta$ values of the
differential samples. The ``top'' result contains
only the top-quark loop, while ``$t+b$'' contains the coherent sum of
the top- and bottom-quark amplitudes. The $t+b$ entries are the samples of
Table~\ref{tab:dist-xsec}, combining \InclusiveTbRunsPerPoint{} independent
integrations with different random-number seeds; the top-only entries combine
\InclusiveTopRunsPerPoint{} at the same integration settings. The quoted
uncertainties are the Monte Carlo integration errors of the combination,
scaled by $\sqrt{\chi^2/\text{ndf}}$ where the runs scatter by more than their
errors.
}
\label{tab:A-inclusive}
\begin{tabular}{c r@{\,$\pm$\,}l r@{\,$\pm$\,}l c}
\hline\hline
$\tan\beta$
& \multicolumn{2}{c}{$\sigma_{\rm top}$ [fb]}
& \multicolumn{2}{c}{$\sigma_{t+b}$ [fb]}
& $\sigma_{t+b}/\sigma_{\rm top}$
\\
\hline
1
& 4000.4 & 1.2
& 4040.7 & 1.2
& 1.010
\\
7
& 81.643 & 0.029
& 258.366 & 0.069
& 3.165
\\
10
& 39.998 & 0.013
& 359.831 & 0.079
& 8.996
\\
15
& 17.7765 & 0.0048
& 688.36 & 0.16
& 38.7
\\
30
& 4.4443 & 0.0012
& 2568.36 & 0.78
& 578
\\
\hline\hline
\end{tabular}
 \end{table}

For the top-only contribution, the numerical results follow the
expected Type-II scaling
$$
\sigma_{\rm top}(A+3j)\propto\cot^2\beta
$$
to within $\TanbScanCotDev\%$ over the whole $\tan\beta$ scan of
Sec.~\ref{subsec:parameter_scans}, at every mass. For example, increasing $\tan\beta$ from 1
to $\AtopTenTanb$ reduces the top-only cross section from about $\AtopOnePb$~pb to
$\AtopTenFb$~fb, while at $\tan\beta=\InclusiveTanbMax$ it is only about
$\AtopThirtyFb$~fb.

The situation changes completely once the bottom-quark loop is
included. At $\tan\beta=1$ the $t+b$ result differs from the top-only
prediction by only about $\AratioOne\%$. The relative importance of the
bottom contribution then grows rapidly: the full result is larger
than the top-only result by factors of approximately $\AratioSeven$,
$\AratioTen$, $\AratioFifteen$ and $\AratioThirty$ for
$\tan\beta=\InclusiveTanbRest$, respectively.
The full cross section is consequently non-monotonic. Among the
explicit points in Table~\ref{tab:A-inclusive}, its smallest value
occurs at $\tan\beta=\AtbMinTanb$, after which the enhancement of the
bottom Yukawa coupling drives the cross section upwards.

For comparison, the $CP$-even scalar $h$ with Standard Model couplings gives
$\sigma_{\rm top}=\HsampleTop\pm\HsampleTopErr$~fb and
$\sigma_{t+b}=\HsampleTb\pm\HsampleTbErr$~fb in the same setup: the bottom loop
changes its rate by $\HsampleRatio\%$, against $\AratioOne\%$ for the
pseudoscalar at $\tan\beta=1$.

\subsection{Partonic composition}
\label{subsec:partonic_results}

An important feature of the complete calculation is the inclusion of
the quark-containing subprocesses in addition to the purely gluonic
channel. Table~\ref{tab:A-subprocesses} shows the decomposition of the
cross section for two representative configurations. The first is the
top-dominated point $\tan\beta=1$ with top loops only, while the second
is the bottom-enhanced point $\tan\beta=\InclusiveTanbMax$ with the coherent
top-plus-bottom amplitude.

\begin{table}[t]
\centering
\caption{
Partonic decomposition of $pp\to A+3j$ at
$m_A=\BenchMass$~GeV. The fractions are normalized to the sum of the six
independently integrated exclusive subprocess classes. Sums over
light-quark flavors and the relevant crossing-related channels are
understood. Each class combines \SubprocTableRunsPerPoint{} independent integrations
with different random-number seeds, at the integration settings of
Table~\ref{tab:A-inclusive}; the uncertainties are as there.
}
\label{tab:A-subprocesses}
\begin{tabular}{l r@{\,$\pm$\,}l r r@{\,$\pm$\,}l r}
\hline\hline
Partonic class
& \multicolumn{3}{c}{$\tan\beta=1$, top}
& \multicolumn{3}{c}{$\tan\beta=30$, $t+b$}
\\
& \multicolumn{2}{c}{$\sigma$ [fb]} & fraction [\%] & \multicolumn{2}{c}{$\sigma$ [fb]} & fraction [\%]
\\
\hline
$gg\to gggA$
& 2057.72 & 0.82
& 51.44
& 1555.1 & 1.0
& 60.53\\
$gg\to q\bar q gA$
& 234.149 & 0.065
& 5.85
& 173.034 & 0.055
& 6.74\\
$gq\to ggqA$
& 1454.00 & 0.38
& 36.35
& 771.13 & 0.23
& 30.02\\
$gq\to q\bar q qA$
& 63.216 & 0.020
& 1.58
& 32.719 & 0.015
& 1.27\\
$q\bar q\to gggA$
& 7.6348 & 0.0026
& 0.19
& 0.090121 & 0.000027
& 0.004\\
$q q\to q q gA$
& 183.465 & 0.036
& 4.59
& 36.9254 & 0.0092
& 1.44\\
\hline
Sum
& 4000.18 & 0.91
& 100.00
& 2569.0 & 1.1
& 100.00\\
\hline\hline
\end{tabular}
 \end{table}

For $\tan\beta=1$, the purely gluonic subprocess
$gg\to gggA$ is the largest single channel but accounts for only
about $\FracGluonicTop\%$ of the exclusive sum. The two $gg$-initiated
classes together contribute about $\FracGGTop\%$, while the two $gq$
classes contribute about $\FracGQTop\%$. The remaining quark-initiated
channels provide approximately $\FracQQTop\%$. Thus a calculation restricted to
$gg\to gggA$ misses almost one half of the complete cross section at
this benchmark.

The composition changes when the bottom-quark loop dominates. At
$\tan\beta=\InclusiveTanbMax$, the $gg$-initiated classes contribute about $\FracGGTb\%$
of the exclusive sum, while the $gq$ channels contribute about
$\FracGQTb\%$. The remaining quark-quark channels account for about
$\FracQQTb\%$. The relative importance of the subprocesses therefore
depends on the heavy-quark structure of the loop amplitude as well
as on the corresponding parton luminosities.

The six classes are integrated independently of the inclusive runs of
Table~\ref{tab:A-inclusive}. Their sums reproduce those inclusive cross sections
to better than $\SubprocSumDevBound\%$, within $\SubprocSumPullMax$
standard deviations of the combined integration error.

\subsection{Mass and $\tan\beta$ dependence}
\label{subsec:parameter_scans}

\begin{figure}[!htbp]
\centering
\includegraphics[width=0.90\textwidth]
{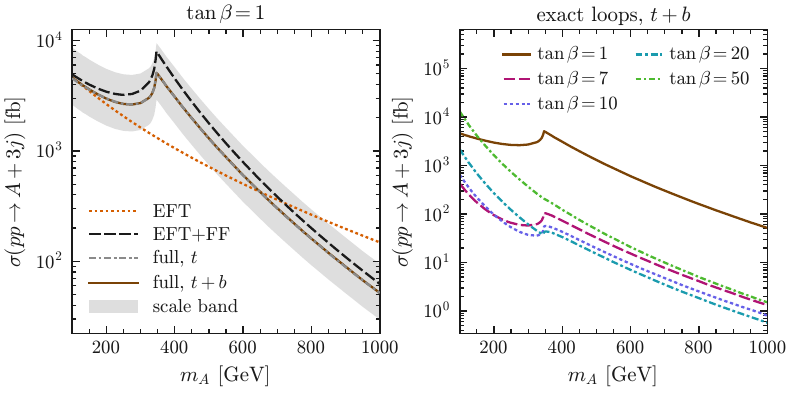}
\caption{
LO cross section for $pp\to A+3j$ at $\sqrt{s}=\Ecm$~TeV as a
function of $m_A$. The left panel compares the four loop treatments at
$\tan\beta=1$: the heavy-top effective theory, the same with the exact
triangle form factor, the exact top loop, and the coherent
top-plus-bottom result. The right panel shows the coherent
top-plus-bottom result for $\tan\beta=\MassScanTanbList$. The top-pair
threshold at $m_A\simeq2m_t$ is prominent in the top-dominated
predictions and becomes progressively less visible as the bottom-loop
contribution increases. The grey band is the
seven-point scale variation of the $t+b$ result at $\tan\beta=1$; over every
point of both scans the variation is the same factor to within a few percent,
+\ScanBandUp\% and $-$\ScanBandDown\%, so it is drawn once. This figure and
Figs.~\ref{fig:A-tanb-scan}--\ref{fig:deta-PHI} were produced by plotting code
written with AI assistance from the stored run outputs; see the
Acknowledgments.
}
\label{fig:A-mass-scan}
\end{figure}

\begin{figure}[!htbp]
\centering
\includegraphics[width=0.90\textwidth]
{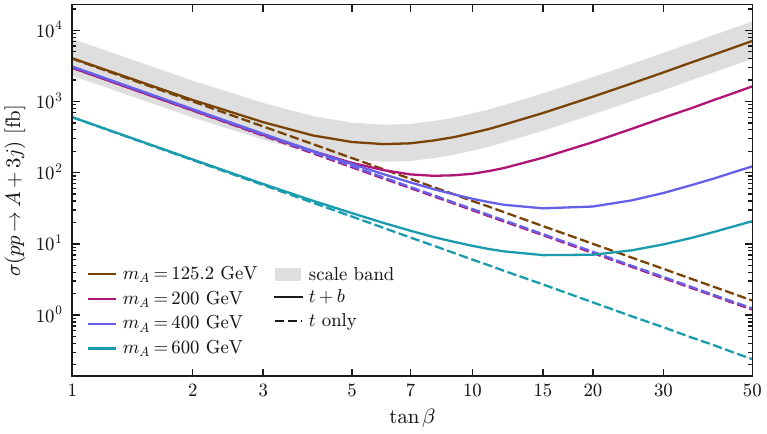}
\caption{
LO cross section for $pp\to A+3j$ at $\sqrt{s}=\Ecm$~TeV as a
function of $\tan\beta$ for $m_A=\TanbScanMasses$~GeV.
Solid curves denote the coherent top-plus-bottom result and dashed
curves the exact top-only prediction; the top-only curves follow the
$\cot^2\beta$ scaling as straight lines on these axes. The grey band is
the seven-point scale variation of the $t+b$ result at the lightest mass, as in
Fig.~\ref{fig:A-mass-scan}.
}
\label{fig:A-tanb-scan}
\end{figure}

The interplay between the heavy-quark masses and the Type-II Yukawa
couplings is illustrated by the mass and $\tan\beta$ scans in
Figs.~\ref{fig:A-mass-scan} and \ref{fig:A-tanb-scan}.

For small $\tan\beta$, where the top-quark loop dominates, the exact
prediction develops the characteristic top-pair threshold structure
around
$$
m_A\simeq2m_t\simeq345~{\rm GeV},
$$
peaking at $m_A=\MassScanPeak$~GeV among the computed points.
The strict heavy-top effective theory does not contain this threshold, and
at $m_A=\MassScanMax$~GeV it exceeds the exact result by a factor
$\MassScanEFTHighRatio$. The form-factor correction restores the
triangle-like threshold behavior, but the resulting EFT+FF prediction still
differs from the full calculation because the box, pentagon and hexagon
amplitudes contain additional momentum dependence: it overshoots the exact
result by a factor $\MassScanFFPeakRatio$ at the threshold peak and
$\MassScanFFHighRatio$ at $m_A=\MassScanMax$~GeV.

The difference between the top-only and $t+b$ curves becomes
progressively larger as $\tan\beta$ increases. At large $\tan\beta$
the bottom-quark loop becomes the dominant contribution and neither
the heavy-top EFT nor a top-based approximation can reproduce the
full result.

The top-only curves decrease monotonically with $\tan\beta$, following
the $\cot^2\beta$ dependence of the pseudoscalar top Yukawa coupling.
Once the bottom loop is included, each curve instead develops a
minimum. Below the minimum the cross section is top dominated, while
above it the $\tan\beta$ enhancement of the bottom Yukawa coupling
drives the cross section upwards. The position of the minimum changes
with $m_A$, reflecting the different mass dependence of the top- and
bottom-loop form factors: among the computed points it lies at
$\tan\beta=\TanbScanMinList$ for $m_A=\TanbScanMasses$~GeV, respectively.

\begin{figure}[!htbp]
\centering
\includegraphics[width=0.90\textwidth]
{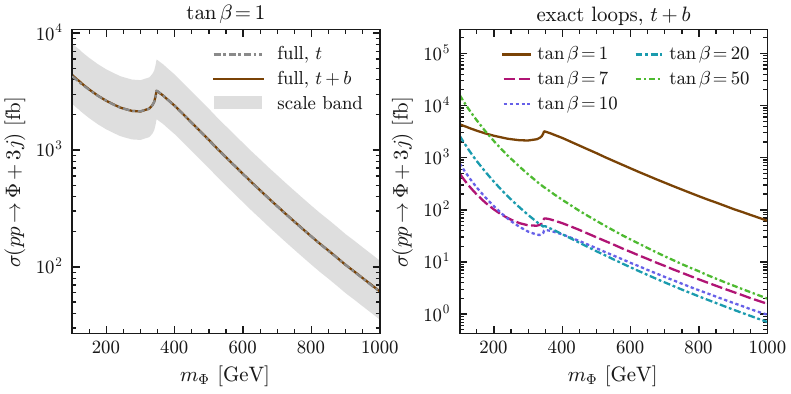}
\caption{
LO cross section for $pp\to \Phi+3j$ at $\sqrt{s}=\Ecm$~TeV as a
function of $m_\Phi$, for the $CP$-mixed state of
Sec.~\ref{subsec:mixed_state}. Left: the exact top-only and top-plus-bottom
results at $\tan\beta=1$, with the seven-point scale variation of the latter.
Right: the top-plus-bottom result for $\tan\beta=\MassScanTanbList$.
}
\label{fig:PHI-mass-scan}
\end{figure}

\begin{figure}[!htbp]
\centering
\includegraphics[width=0.90\textwidth]
{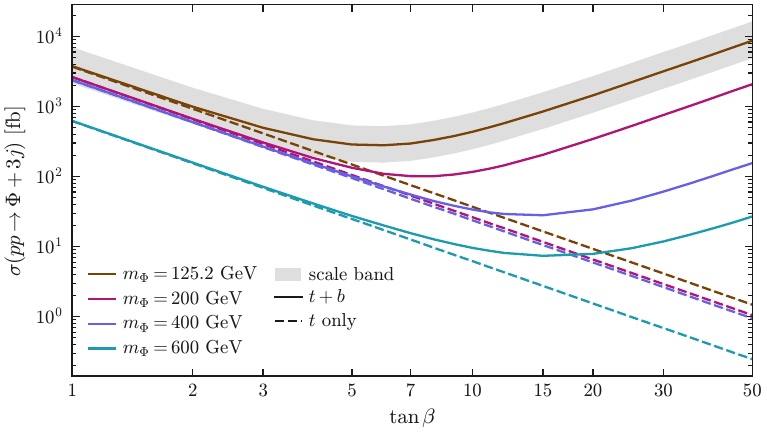}
\caption{
LO cross section for $pp\to\Phi+3j$ at $\sqrt{s}=\Ecm$~TeV as a function of
$\tan\beta$ for $m_\Phi=\TanbScanMasses$~GeV, as in
Fig.~\ref{fig:A-tanb-scan}: solid curves top-plus-bottom, dashed top-only,
and the grey band the scale variation at the lightest mass.
}
\label{fig:PHI-tanb-scan}
\end{figure}

The same scans for the $CP$-mixed state $\Phi$ of
Sec.~\ref{subsec:mixed_state} are shown in Figs.~\ref{fig:PHI-mass-scan}
and~\ref{fig:PHI-tanb-scan}, with the exact top and top-plus-bottom loops.
They follow the pseudoscalar closely. The top-only cross section again
scales as $\cot^2\beta$, to within $\PhiTanbScanCotDev\%$, and the
top-plus-bottom curves have their minimum at the same points,
$\tan\beta=\PhiTanbScanMinList$ for $m_\Phi=\TanbScanMasses$~GeV. At
$\tan\beta=1$ the bottom loop changes the $\Phi$ rate by at most
$\PhiMassScanBotMax\%$ over the whole mass range. The top-pair threshold
is clearly weaker than for $A$: at $\tan\beta=1$ the cross section rises
from the minimum below $2m_t$ to the threshold peak by a factor
$\ThresholdRatioPhi$, against $\ThresholdRatioA$ for the pseudoscalar. The
weaker threshold enhancement is consistent with the different threshold
behaviour of scalar and pseudoscalar top-quark loops: the scalar current
couples to a $P$-wave $t\bar t$ configuration, whereas the pseudoscalar
current couples to an $S$-wave configuration~\cite{Spira1998a}.

\subsection{Finite-mass effects in differential distributions}
\label{subsec:finite_mass_distributions}

We next examine differential distributions for the complete
$A+3$-jet process at $m_A=\BenchMass$~GeV. The comparison with the
form-factor-improved effective theory is made at $\tan\beta=1$. In
addition, the full top-plus-bottom prediction is shown for
$\tan\beta=\DistTanbList$. Every sample is the bin-by-bin combination of
\DistRunsPerPoint{} independent integrations with different random-number
seeds, run on two different computing clusters.

All distributions in this subsection are normalized to unit area, so
that the shapes can be compared directly without the overall rates,
which differ by more than an order of magnitude across the
$\tan\beta$ range, dominating the comparison. The rates themselves are
given in Table~\ref{tab:dist-xsec}: among the full top-plus-bottom
samples the total cross section varies by a factor of about
$\DistXsecSpread$, with its smallest value at
$\tan\beta=\DistXsecMinTanb$, and the EFT+FF prediction exceeds the
exact result at $\tan\beta=1$ by a factor $\DistEffOverFull$.

\begin{table}[tbp]
\centering
\caption{Total cross sections of the samples shown in
Figs.~\ref{fig:ptA}--\ref{fig:deta-PHI}, at $\sqrt{s}=\Ecm$~TeV and
$m_A=m_\Phi=m_h=\BenchMass$~GeV. The uncertainties are the Monte Carlo integration
errors of the combined runs, scaled by $\sqrt{\chi^2/\text{ndf}}$ where the
runs scatter by more than their errors (App.~\ref{app:numerics}). These
cross sections give the normalizations before any observable-specific
selection. The $\Phi_2$ distributions are instead normalized to the
cross section passing Eq.~\eqref{eq:phi2_sel}.}
\label{tab:dist-xsec}
\begin{tabular}{l l c r@{\,$\pm$\,}l c c c}
\hline\hline
state & sample & $\tan\beta$ & \multicolumn{2}{c}{$\sigma$ [fb]} & $\delta\sigma/\sigma$ [\%] & $n$ & $\sigma/\sigma_{A,\,t+b,\,\tan\beta=1}$ \\
\hline
$A$ & EFT+FF & $1$ & 4401.59 & 0.61 & 0.014 & 12 & 1.0893 \\
$A$ & full $t+b$ & $1$ & 4040.70 & 1.20 & 0.030 & 12 & 1.0000 \\
$A$ & full $t+b$ & $7$ & 258.37 & 0.07 & 0.027 & 12 & 0.0639 \\
$A$ & full $t+b$ & $10$ & 359.83 & 0.08 & 0.022 & 12 & 0.0891 \\
$A$ & full $t+b$ & $15$ & 688.36 & 0.16 & 0.023 & 12 & 0.1704 \\
$A$ & full $t+b$ & $30$ & 2568.36 & 0.78 & 0.030 & 12 & 0.6356 \\
$\Phi$ & full $t+b$ & $1$ & 3778.07 & 0.90 & 0.024 & 12 & 0.9350 \\
$\Phi$ & full $t+b$ & $7$ & 299.14 & 0.07 & 0.023 & 12 & 0.0740 \\
$\Phi$ & full $t+b$ & $10$ & 437.51 & 0.10 & 0.022 & 12 & 0.1083 \\
$\Phi$ & full $t+b$ & $15$ & 851.20 & 0.19 & 0.022 & 12 & 0.2107 \\
$\Phi$ & full $t+b$ & $30$ & 3184.60 & 0.86 & 0.027 & 12 & 0.7881 \\
$h$ & EFT+FF & $1$ & 2053.94 & 0.30 & 0.015 & 12 & 0.5083 \\
$h$ & full $t+b$ & $1$ & 1982.17 & 0.35 & 0.018 & 12 & 0.4906 \\
\hline\hline
\end{tabular}
 \end{table}

\begin{figure}[!htbp]
\centering
\includegraphics[width=0.92\textwidth]
{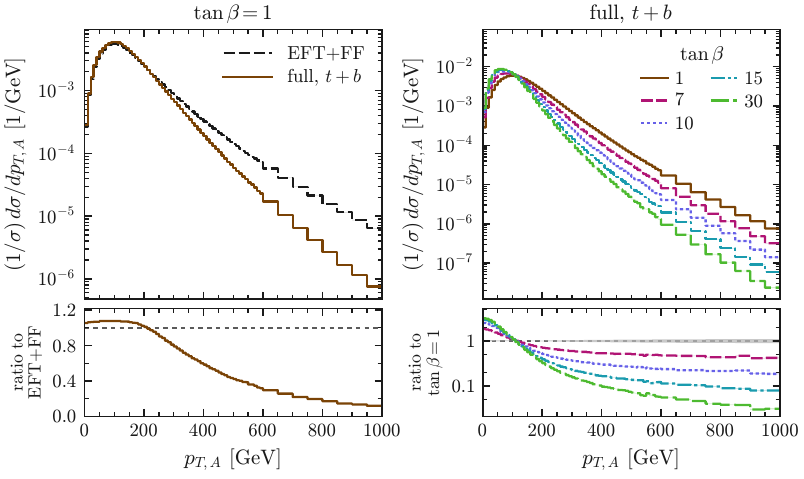}
\caption{
Normalized distribution in the pseudoscalar transverse momentum
$p_{T,A}$ for $pp\to A+3j$ at $\sqrt{s}=\Ecm$~TeV and
$m_A=\BenchMass$~GeV. The left panel compares EFT+FF with the exact
top-plus-bottom result at $\tan\beta=1$. The right panel shows the
exact top-plus-bottom prediction for several values of $\tan\beta$.
The lower panels show ratios to EFT+FF and to the $\tan\beta=1$
prediction, respectively. \RebinPtHiggsSentence{} The corresponding
total cross sections are given in Table~\ref{tab:dist-xsec}.
In the lower right panel the grey band is the seven-point $(\mu_R,\mu_F)$ variation of the normalized $\tan\beta=1$ prediction, each scale point normalized on its own: the uncertainty on the shape, which the $\tan\beta$ dependence is to be read against. Under the same variation every rate moves up by \COneRateBandUp\% and down by \COneRateBandDown\%, and the ratio of EFT+FF to the full result, both at a common scale, by at most $\COneSameScaleSpread\%$.
}
\label{fig:ptA}
\end{figure}

The Higgs transverse momentum provides the clearest illustration of
the breakdown of the effective description, shown in
Fig.~\ref{fig:ptA}.
At low transverse momentum, the EFT+FF and full predictions are close.
Their difference grows rapidly as $p_{T,A}$ approaches and exceeds
the top-quark mass. In the high-$p_T$ tail the form-factor-improved
effective theory is substantially harder than the exact result. The
triangle form factor therefore does not reproduce the suppression
associated with resolving the full higher-point heavy-quark loop.
The dependence on $\tan\beta$ is also strongly kinematic: beyond the
change in overall rate recorded in Table~\ref{tab:dist-xsec}, the
shape of the spectrum itself changes. The bottom-enhanced predictions
are relatively more concentrated at low transverse momentum, while the
hard tail is suppressed relative to the top-dominated $\tan\beta=1$
result.

\begin{figure}[!htbp]
\centering
\includegraphics[width=0.92\textwidth]
{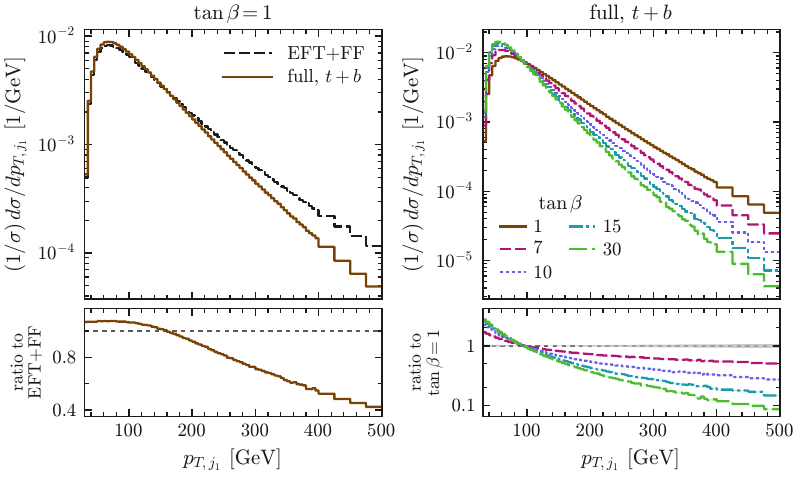}
\caption{
Normalized distribution in the transverse momentum of the hardest jet,
$p_{T,j_1}$. The left panel compares EFT+FF with the full top-plus-bottom
result at $\tan\beta=1$, while the right panel shows the full prediction for
several values of $\tan\beta$. \RebinPtjoneSentence{}
The grey band is the scale variation of the $\tan\beta=1$ shape, as in
Fig.~\ref{fig:ptA}.
}
\label{fig:ptj1}
\end{figure}

\begin{figure}[!htbp]
\centering
\includegraphics[width=0.92\textwidth]
{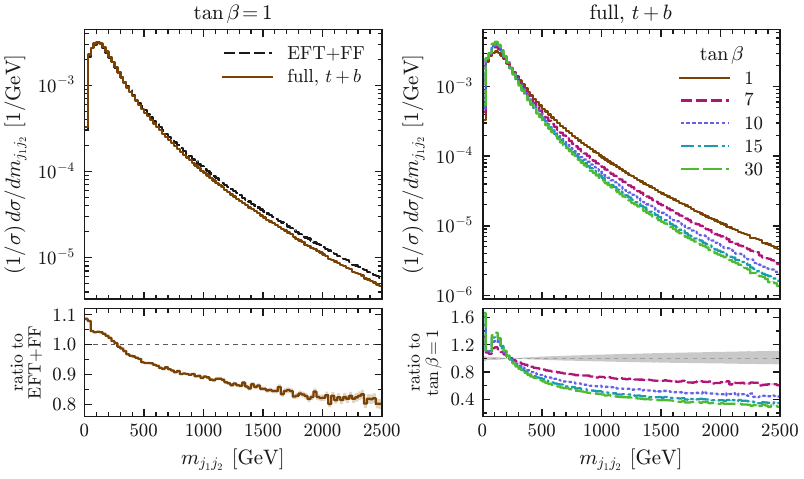}
\caption{
Normalized distribution in the invariant mass $m_{j_1j_2}$ of the
two leading jets. The left panel compares EFT+FF with the full
top-plus-bottom prediction at $\tan\beta=1$, while the right panel
shows the dependence of the normalized full result on $\tan\beta$.
\RebinMjjSentence{}
The grey band is the scale variation of the $\tan\beta=1$ shape, as in Fig.~\ref{fig:ptA}.
}
\label{fig:mjj-normalized}
\end{figure}

\begin{figure}[!htbp]
\centering
\includegraphics[width=0.92\textwidth]
{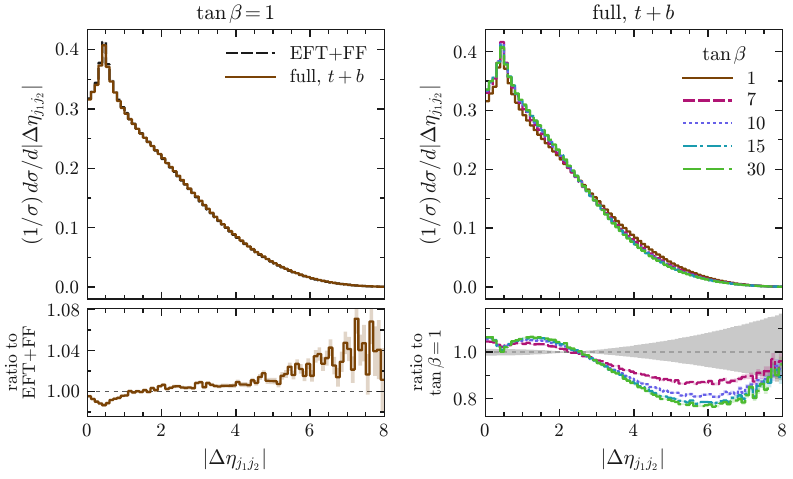}
\caption{
Normalized distribution in
$|\Delta\eta_{j_1j_2}|=|\eta_{j_1}-\eta_{j_2}|$ for
$pp\to A+3j$. The left panel compares EFT+FF with the full
top-plus-bottom result at $\tan\beta=1$ and the right panel shows the
full prediction for different values of $\tan\beta$.
\RebinDetaSentence{}
The grey band is the scale variation of the $\tan\beta=1$ shape, as in Fig.~\ref{fig:ptA}.
}
\label{fig:deta-normalized}
\end{figure}

The hardest jet recoils against the boson and the other jets, and its
transverse momentum, Fig.~\ref{fig:ptj1}, shows the same breakdown of the
effective description, setting in at a somewhat lower transverse momentum
than for $p_{T,A}$. The bottom-enhanced predictions are again softer.
A similar effect is visible in the invariant mass of the two leading
jets. To isolate the shape dependence, we show the normalized
distribution in Fig.~\ref{fig:mjj-normalized}.
The EFT+FF description becomes progressively harder than the exact
result as $m_{j_1j_2}$ increases. The normalized $\tan\beta$ scan also
shows that bottom-enhanced production is relatively depleted in the
large-$m_{j_1j_2}$ region. These distortions are established for the
inclusive selection used here; their size under VBF selections is not
determined by this comparison.
The rapidity separation of the two leading jets is shown in
Fig.~\ref{fig:deta-normalized}.
In contrast to $p_{T,A}$ and $m_{j_1j_2}$, the normalized
$|\Delta\eta_{j_1j_2}|$ distribution is rather insensitive to the
heavy-quark treatment over most of the populated phase space.
The $\tan\beta$ dependence is larger: the bottom-enhanced samples are
slightly more populated at small separations and depleted, by up to about a
fifth, at large ones, well outside the scale band. This is the same
softening seen in $p_{T,A}$ and $m_{j_1j_2}$. The large fluctuations at
the end of the plotted range arise where the differential cross
section has become very small.

\subsection{The third jet}
\label{subsec:thirdjet}

\begin{figure}[!htbp]
\centering
\includegraphics[width=0.92\textwidth]
{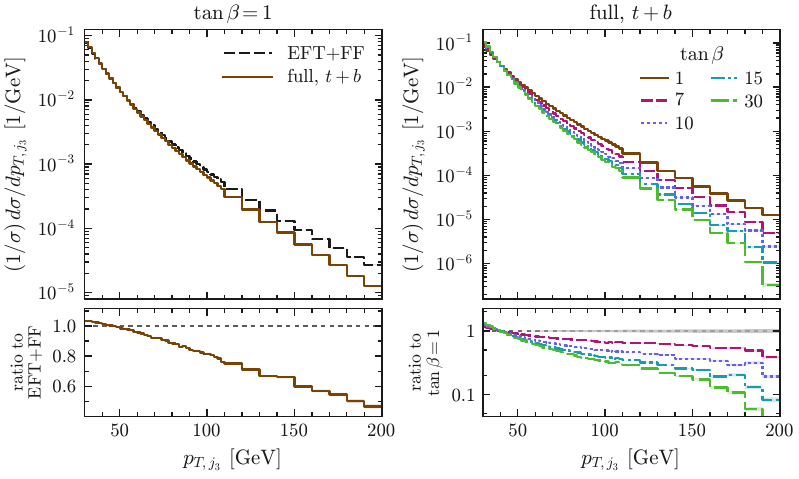}
\caption{
Normalized distribution in the transverse momentum of the third
jet, $p_{T,j_3}$. The left panel compares EFT+FF with the full
top-plus-bottom result at $\tan\beta=1$, while the right panel shows
the full prediction for several values of $\tan\beta$.
\RebinPtjthreeSentence{} The corresponding total cross sections are given in
Table~\ref{tab:dist-xsec}.
The grey band is the scale variation of the $\tan\beta=1$ shape, as in Fig.~\ref{fig:ptA}.
}
\label{fig:ptj3}
\end{figure}

The third jet directly probes the additional hard QCD radiation
present in the three-jet process. Figure~\ref{fig:ptj3} shows its
normalized transverse-momentum distribution.
The difference between EFT+FF and the exact calculation increases
with $p_{T,j_3}$. The effective prediction is therefore increasingly
unable to describe hard third-jet radiation. The $\tan\beta$
dependence also changes the shape of the spectrum, with the
bottom-enhanced results becoming softer as $\tan\beta$ is increased.

\begin{figure}[!htbp]
\centering
\includegraphics[width=0.92\textwidth]
{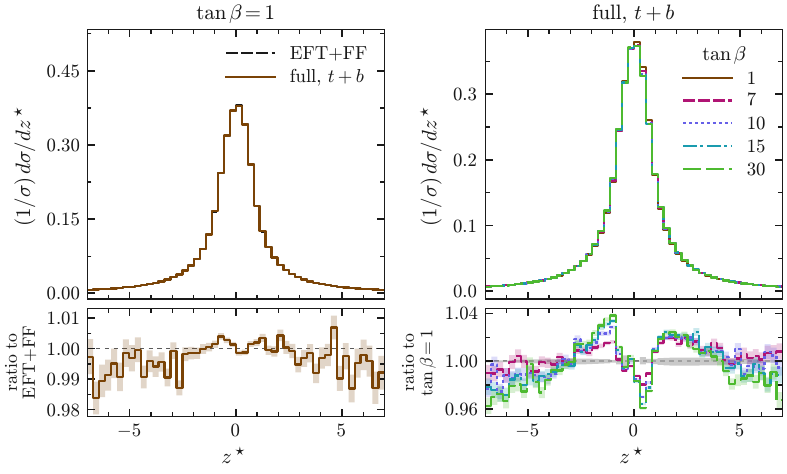}
\caption{
Normalized third-jet centrality distribution $z^\star$ for
$pp\to A+3j$. The left panel compares EFT+FF with the full
top-plus-bottom prediction at $\tan\beta=1$ and the right panel shows
the full result for different values of $\tan\beta$.
\RebinZstarSentence{}
The grey band is the scale variation of the $\tan\beta=1$ shape, as in Fig.~\ref{fig:ptA}.
}
\label{fig:zstar}
\end{figure}

We also study the position of the third jet relative to the two
leading jets using
$$
z^\star=
\frac{
\eta_{j_3}-\frac{1}{2}(\eta_{j_1}+\eta_{j_2})
}{
|\eta_{j_1}-\eta_{j_2}|
}.
$$
The normalized distribution is shown in Fig.~\ref{fig:zstar}.
The $z^\star$ distribution is sharply concentrated around
$z^\star=0$, corresponding to a third jet close to the midpoint of
the two leading jets in pseudorapidity. Unlike the transverse-momentum
observables, its normalized shape is remarkably stable. The EFT+FF
and full predictions are nearly identical after normalization, and
the dependence on $\tan\beta$ is small. The large bottom-loop effects
seen in the total rate therefore do not translate into a comparable
distortion of this centrality observable.

\subsection{Azimuthal correlations}
\label{subsec:azimuthal_results}

Azimuthal correlations between the jets are sensitive to the tensor
structure of the Higgs--gluon interaction. In $\Phi+2$-jet production
the standard observable is the signed azimuthal-angle difference
$\phi_{jj}$ between the two tagging jets. With a third hard jet in the
event that two-jet variable is no longer the natural choice, and we
use instead a multi-jet observable $\Phi_2$ following
Ref.~\cite{Andersen:2010zx}, which was constructed to retain the
$CP$ sensitivity of $\phi_{jj}$ in the presence of additional hard
radiation.

The jets are divided into the two hemispheres defined by the rapidity
$y_\Phi$ of the Higgs boson, and the jet momenta in each hemisphere
are summed,
\begin{equation}
\label{eq:phi2_sums}
q_a=\sum_{j\,:\,y_j<y_\Phi}p_j,
\qquad
q_b=\sum_{j\,:\,y_j>y_\Phi}p_j.
\end{equation}
The observable is the azimuthal angle between the transverse parts of
these two vectors,
\begin{equation}
\label{eq:phi2_def}
\Phi_2=\angle\left(\mathbf{q}_{a\perp},\mathbf{q}_{b\perp}\right),
\qquad
\cos\Phi_2
=\frac{\mathbf{q}_{a\perp}\cdot\mathbf{q}_{b\perp}}
      {\left|\mathbf{q}_{a\perp}\right|\,
       \left|\mathbf{q}_{b\perp}\right|},
\end{equation}
with the sign of
\begin{equation}
\label{eq:phi2_sign}
\left(\mathbf{q}_{a\perp}\times\mathbf{q}_{b\perp}\right)\cdot\hat{z},
\end{equation}
where $\hat z$ is the unit vector along the beam axis whose sign is that of
the $z$ component of $q_a-q_b$~\cite{Andersen:2010zx}. $\Phi_2$ then covers
the full range $-180^\circ\leq\Phi_2\leq180^\circ$, and its distribution is
symmetric about $\Phi_2=0$ for a $CP$ eigenstate.

As in Ref.~\cite{Andersen:2010zx}, an event enters the distribution only if
at least one jet lies on each side of the boson in rapidity,
\begin{equation}
\label{eq:phi2_sel}
\exists\, j_a:\ y_{j_a}<y_\Phi,
\qquad
\exists\, j_b:\ y_{j_b}>y_\Phi,
\end{equation}
without a further minimum rapidity separation between the jets and the
boson. The requirement is physical rather than technical: the factorization
of the amplitude at large rapidity separations, which ties the azimuthal
angle between $q_a$ and $q_b$ to the $CP$ structure of the effective
Higgs--gluon vertex, holds for a boson inside the rapidity interval spanned
by the partons~\cite{Andersen:2010zx}, and events with all the radiation on
one side of the boson do not carry that correlation. In an exclusive
three-jet sample such events are frequent: the selection of
Eq.~\eqref{eq:phi2_sel} keeps \PhiTwoKeepAAZ\% of the cross section.
Each $\Phi_2$ distribution is normalized to unit area using its own
cross section $\sigma_{\rm sel}$ after the selection in
Eq.~\eqref{eq:phi2_sel}, rather than the inclusive cross section in
Table~\ref{tab:dist-xsec}.

We characterize the modulation by the relative amplitude $A_2$ of its second
harmonic. We fit the distribution with a Fourier series up to the fourth
harmonic:
\begin{equation}
\label{eq:a2}
\frac{d\sigma}{d\Phi_2}\propto1+\sum_{n=1}^{4}\left(c_n\cos n\Phi_2+s_n\sin n\Phi_2\right),
\qquad
A_2=\sqrt{c_2^2+s_2^2},
\end{equation}
At $\tan\beta=1$, with the exact top-plus-bottom loops, the fit gives
$A_2=\PhiTwoAmpAAAZ$ for the pseudoscalar $A$ and $A_2=\PhiTwoAmpHAAZ$ for
the $CP$-even $h$ with Standard Model couplings, whose $\cos2\Phi_2$ term
has the opposite sign. We focus on $A_2$ because the characteristic
$CP$-sensitive azimuthal correlation appears in the second harmonic; this
does not imply that the first harmonic vanishes or is numerically negligible.

\begin{figure}[!htbp]
\centering
\includegraphics[width=0.92\textwidth]
{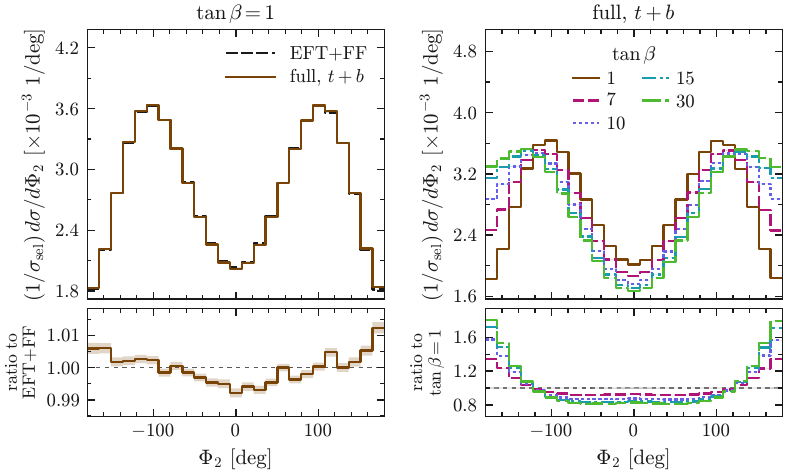}
\caption{
Normalized distribution in the signed multi-jet azimuthal observable
$\Phi_{2}$, Eqs.~\eqref{eq:phi2_sums}--\eqref{eq:phi2_sel}, for
$pp\to A+3j$. The left panel compares EFT+FF with
the full top-plus-bottom result at $\tan\beta=1$, while the right
panel shows the dependence of the full result on $\tan\beta$.
Since the modulation is smooth across the full range,
\RebinPhitwojjFactor{} native bins are combined everywhere.
The grey band is the scale variation of the $\tan\beta=1$ shape, as in Fig.~\ref{fig:ptA}.
}
\label{fig:phi2jj}
\end{figure}

The normalized $\Phi_2$ distribution is shown in
Fig.~\ref{fig:phi2jj}.
At $\tan\beta=1$, the normalized EFT+FF and exact distributions are
very similar. This shows that, in the top-dominated region, the
azimuthal shape is considerably more stable against finite-mass
effects than the hard transverse-momentum distributions. The
situation changes as the bottom loop becomes important. Increasing
$\tan\beta$ produces a visible redistribution of events across
$\Phi_{2}$, demonstrating that the bottom-quark contribution can
modify the angular correlation even though the state remains
$CP$ odd.

\subsection{The $CP$-mixed state $\Phi$}
\label{subsec:phi_distributions}

The same observables are available for the $CP$-mixed state $\Phi$ at
$m_\Phi=\BenchMass$~GeV, from the same set of runs as
Sec.~\ref{subsec:finite_mass_distributions}, with the couplings of
Sec.~\ref{subsec:mixed_state}. We have not computed an EFT+FF prediction for
the mixed state, so only the $\tan\beta$ dependence of the exact
top-plus-bottom result is shown. As a check of the mixed-state implementation, the
interference of the two components integrates to zero in the total rate: at
$\tan\beta=1$ the $\Phi$ cross section of Table~\ref{tab:dist-xsec} equals
$c_H^2\sigma(h)+c_A^2\sigma(A)$ from the pure-state samples to a ratio of
$\PhiSumRuleRatio\pm\PhiSumRuleRatioErr$, i.e.\ within $\PhiSumRulePull$
standard deviations. The smaller $\Phi$ rate relative to $A$ at
$\tan\beta=1$ reflects the chosen scalar and pseudoscalar coupling weights.

\begin{figure}[!htbp]
\centering
\includegraphics[width=0.5\textwidth]
{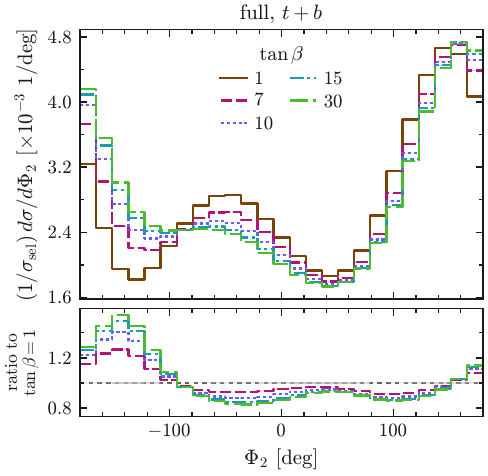}
\caption{
Normalized distribution in the signed multi-jet azimuthal observable
$\Phi_{2}$ for $pp\to\Phi+3j$, for the available values of
$\tan\beta$. As in Fig.~\ref{fig:phi2jj},
\RebinPhitwojjFactor{} native bins are combined everywhere.
The lower panel shows the ratio to the $\tan\beta=1$ prediction; the grey
band is the scale variation of the $\tan\beta=1$ shape, as in
Fig.~\ref{fig:ptA}.
Total cross sections are given in
Table~\ref{tab:dist-xsec}.
}
\label{fig:phi2jj-PHI}
\end{figure}

\begin{figure}[!htbp]
\centering
\begin{minipage}[t]{0.48\textwidth}
\centering
\includegraphics[width=\textwidth]
{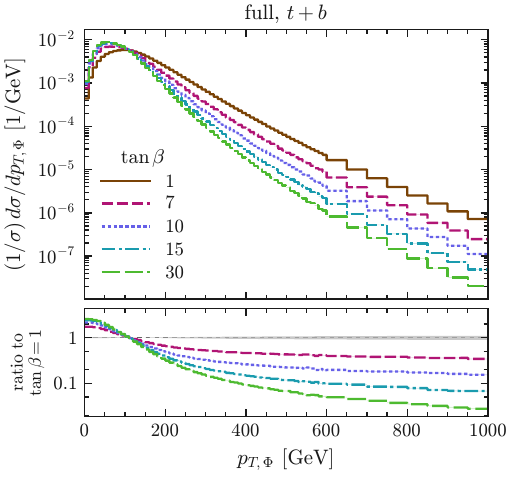}
\caption{
Normalized distribution in the transverse momentum of the $CP$-mixed
state, $p_{T,\Phi}$, for $pp\to\Phi+3j$. \RebinPtHiggsSentence{}
The lower panel shows the ratio to the $\tan\beta=1$ prediction; the grey
band is the scale variation of the $\tan\beta=1$ shape, as in
Fig.~\ref{fig:ptA}.
}
\label{fig:ptPhi-PHI}
\end{minipage}\hfill
\begin{minipage}[t]{0.48\textwidth}
\centering
\includegraphics[width=\textwidth]
{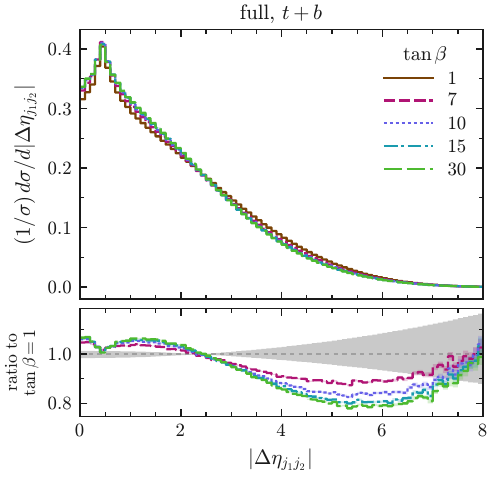}
\caption{
Normalized distribution in $|\Delta\eta_{j_1j_2}|$ for
$pp\to\Phi+3j$. \RebinDetaSentence{}
The lower panel and the grey band are as in Fig.~\ref{fig:ptPhi-PHI}.
}
\label{fig:deta-PHI}
\end{minipage}
\end{figure}

Unlike the $CP$ eigenstates, $\Phi$ has a $\Phi_2$ distribution that is not
symmetric under $\Phi_2\to-\Phi_2$, Fig.~\ref{fig:phi2jj-PHI}: the
$CP$-even--$CP$-odd interference adds a $\sin2\Phi_2$ term. The simplest
measure of it needs no fit, the counting asymmetry
\begin{equation}
\label{eq:asin}
A_{\sin}=\frac{\sigma_{\rm sel}(\sin2\Phi_2>0)-\sigma_{\rm sel}(\sin2\Phi_2<0)}{\sigma_{\rm sel}},
\end{equation}
where $\sigma_{\rm sel}$ is the cross section passing the $\Phi_2$
selection of Eq.~\eqref{eq:phi2_sel}. The asymmetry vanishes for any
$CP$ eigenstate and cannot be generated by a $CP$-blind kinematic weight
that is even in $\Phi_2$. For the $A$ and $h$ samples it
is compatible with zero, within $\PhiAsymNullMax$ standard deviations of the
integration error. For $\Phi$ it is $\PhiAsymLo\%$ at $\tan\beta=1$ and
$\PhiAsymHi\%$ at $\tan\beta=\InclusiveTanbMax$, its sign being that of the
interference term in the convention of Eq.~\eqref{eq:phi2_sign}: it decreases
in magnitude with $\tan\beta$ but remains well resolved where the bottom loop
dominates. The
$p_{T,\Phi}$ and $|\Delta\eta_{j_1j_2}|$ distributions,
Figs.~\ref{fig:ptPhi-PHI} and~\ref{fig:deta-PHI}, follow the pattern of
the pseudoscalar: the spectrum softens as the bottom loop takes over.

\FloatBarrier
\section{Conclusions}
\label{sec:summ}

We have studied Higgs boson production in association with three jets
through gluon fusion, including the complete set of partonic
subprocesses. The calculation is loop induced already at leading
order and contains box, pentagon and hexagon contributions. The full
top- and bottom-quark mass dependence is kept in the amplitudes, and
the top and bottom contributions are combined coherently.
A main aim of this work was to go beyond the purely gluonic
subprocess $gg\to ggg\Phi$. We find that the quark-containing
channels are numerically important. For a pseudoscalar Higgs boson
with $m_A=\BenchMass$~GeV and $\tan\beta=1$, the purely gluonic channel
$gg\to gggA$ contributes only about $\FracGluonicTop\%$ of the complete cross
section with top-quark loops. The $gq$-initiated channels alone provide about
$\FracGQTop\%$ of the exclusive sum. This shows that the full set of partonic
channels is needed for a reliable prediction of the $A+3$-jet rate.

The bottom-quark contribution becomes increasingly important when
$\tan\beta$ is increased. For $\tan\beta=1$, including the bottom
loop changes the total pseudoscalar cross section by only about
$\AratioOne\%$. At larger $\tan\beta$, the top contribution is suppressed by
the $\cot\beta$ dependence of the top Yukawa coupling, while the
bottom contribution is enhanced by $\tan\beta$. As a result, the
cross section first decreases and then rises again. At
$\tan\beta=\InclusiveTanbMax$, for example, the top-only prediction is
about $\AtopThirtyFb$~fb, while the full top-plus-bottom result is
about $\AtbThirtyPb$~pb, larger by a factor of roughly
$\AratioThirty$. The exact bottom-quark contribution is therefore essential in
this region.

We have also compared the full mass-dependent calculation with the
heavy-top effective theory and with the form-factor-improved
approximation. The differences become particularly clear in hard
kinematic regions. The form-factor-improved prediction gives a much
harder Higgs transverse-momentum spectrum than the full calculation
once $p_{T,A}$ becomes comparable to or larger than the top-quark
mass. A similar behavior is seen in the transverse momenta of the
hardest and of the third jet and in the large-$m_{j_1j_2}$ region. The triangle form
factor therefore does not reproduce the full momentum dependence of
the box, pentagon and hexagon amplitudes.

Not all observables are equally sensitive to the heavy-quark
structure. The normalized $|\Delta\eta_{j_1j_2}|$ and $z^\star$
distributions are almost unchanged between EFT+FF and the exact result,
and $z^\star$ also hardly depends on $\tan\beta$. In contrast, the
normalized $m_{j_1j_2}$ and transverse momentum distributions, and the
large-separation tail of $|\Delta\eta_{j_1j_2}|$, show a clear dependence
on $\tan\beta$.
The third-jet transverse momentum is particularly useful because it
directly probes the additional hard radiation present in the
three-jet process.

We have also studied the multi-jet azimuthal correlation $\Phi_2$,
built from the jet momenta on the two sides of the Higgs boson in
rapidity. For $\tan\beta=1$, the normalized angular distribution is
described rather well by the form-factor-improved effective theory.
When the bottom-quark loop becomes important, however, the shape of
the distribution changes. This shows that heavy-quark mass effects
can influence not only the total rate but also angular observables
used to study the structure of the Higgs--gluon interaction.

The results presented here show that both the complete set of
partonic subprocesses and the exact heavy-quark mass dependence are
important for a reliable description of Higgs-plus-three-jet
production.
The quark-initiated
channels make up a sizeable fraction of the cross section, while
finite top- and bottom-quark mass effects become large in high-energy
tails and at large $\tan\beta$. These effects should therefore be
included when Higgs-plus-three-jet production is used for precision
studies of gluon-fusion Higgs production and for analyses based on
jet kinematics and angular correlations.

Finally, we have compared the $CP$-even state $h$, the $CP$-odd state $A$
and the $CP$-mixed state $\Phi$ in the same observable. The two
eigenstates are modulated in $\cos2\Phi_2$ with opposite signs; at
$\tan\beta=1$ the relative amplitudes of that modulation,
Eq.~\eqref{eq:a2}, are $\PhiTwoAmpHAAZ$ and $\PhiTwoAmpAAAZ$, so the
pseudoscalar is the more diluted. The mixed state is distinguished from both by a $\sin2\Phi_2$ term,
which no $CP$ eigenstate can produce: its counting asymmetry,
Eq.~\eqref{eq:asin}, is $\PhiAsymLo\%$ at $\tan\beta=1$, and it remains
clearly non-zero at $\tan\beta=\InclusiveTanbMax$, where the bottom loop
dominates. The
multi-jet $\Phi_2$ observable therefore separates all three cases in
Higgs-plus-three-jet production, and the asymmetry offers a fit-free test of
$CP$ violation in the effective Higgs--gluon coupling.

\begin{acknowledgments}

It is a pleasure to thank Dieter Zeppenfeld for the fruitful discussions
during the development of this project.

A. Bhattacharya, F. Campanario, and I. Rosario are supported by the
Generalitat Valenciana, the Spanish government, and ERDF funds from the
European Commission ``NextGenerationEU/PRTR'' (CNS2022-136165,
PID2023-151418NB-I00, MCIN/AEI/10.13039/501100011033).
This work used IFIC's GLUON computing infrastructure, supported by
MCIU/AEI and ERDF under project EQC2019-006074-P.

Tunde Kushimo acknowledges support from a Graduate Teaching Assistantship provided by 
the Division of Physics within the Department of Mathematics, Statistics, and Physics at Wichita State University. 
This research made use of the BeoShock high-performance computing resources at Wichita State University. 
The authors thank Wichita State University Information Technology Services (ITS), 
the Office of Research, and Academic Affairs for their continued support of local high-performance computing infrastructure. 
The authors also acknowledge the valuable system administration and operational support provided through 
Wichita State University's partnership with the BeoCat High-Performance Computing Group at Kansas State University 
and thank Dan Andresen for his leadership and support in enabling these computational resources.

\emph{Use of AI tools.} Substantial parts of this work were carried out with
the AI coding agent Claude Code (Anthropic), with the models Claude Opus~5
and Claude Opus~5.5, between September and October 2026. Directed by the
authors, the tool (i)~wrote the extensions of VBFNLO that compute the
seven-point scale variation by reweighting and fill the $\Phi_2$ observable
of Ref.~\cite{Andersen:2010zx}; (ii)~generated the input cards, submitted and
monitored the batch runs, and wrote the analysis code that combines the runs
and produces every table, figure and quoted number of
Sec.~\ref{sec:results} and Appendix~\ref{app:numerics} from the stored run
outputs; (iii)~carried out cross-checks of the results; and (iv)~assisted with
the literature search and with drafting and revising the text, including
explanations of the results. The one-loop amplitudes and their implementation
in VBFNLO predate the use of these tools. The authors specified the
calculations, the runs and the checks, and verified the output: every number
in Sec.~\ref{sec:results} and Appendix~\ref{app:numerics} is generated from the
run outputs and checked automatically against the text; independent
integrations performed on two computing clusters with separately built
executables agree, as described in Appendix~\ref{app:numerics}; the $\Phi_2$
implementation was compared with an independent MadGraph5\_aMC@NLO
calculation in the heavy-top limit; and the authors reviewed the code changes
and the text. The authors take full responsibility for the content of this
paper.
\end{acknowledgments}

\appendix

\section{Numerical setup, parameter grids and integration precision}
\label{app:numerics}

This appendix collects the inputs behind the results of
Sec.~\ref{sec:results} and the integration quality achieved. Every entry is
parsed from the input cards and logs of the runs themselves rather than
transcribed by hand.

Table~\ref{tab:setup} gives the complete numerical setup -- masses, couplings,
cuts, scales and the parton distributions -- read from the cards of the runs
that produced the results in this paper and checked to agree across all of
them. Table~\ref{tab:grids} lists the parameter grid each campaign covered.

\begin{table}[htbp]
\centering
\caption{The numerical setup, as read from the input cards of the runs
themselves.}
\label{tab:setup}
\begin{tabular}{ll}
\hline\hline
parameter & value \\
\hline
$\sqrt{s}$ & $\Ecm$~TeV \\
PDF set & \texttt{\PDFset}, member \PDFmember \\
jet algorithm & \JetAlgo, $R=\JetR$ \\
jet cuts & $p_{T,j}>\JetPtMin$~GeV, $|y_j|<\JetYMax$ \\
scales & $\mu_F=\mu_R=H_T/\ScaleDenom$ \\
$\alpha_s$ & $\alpha_s(\mu_R)$ of the PDF set, in all five powers \\
$m_t$ & $\Mtop$~GeV \\
$m_b$ & $\Mbot$~GeV \\
$M_W$ & $\Mw$~GeV \\
$M_Z$ & $\Mz$~GeV \\
$G_F$ & $\GFmantissa\times10^{\GFexponent}$~GeV$^{-2}$ \\
\hline\hline
\end{tabular}
 \end{table}

\begin{table}[htbp]
\centering
\caption{Parameter grids covered by the campaigns entering this paper.}
\label{tab:grids}
\begin{tabular}{lll}
\hline\hline
campaign & states & grid \\
\hline
mass scan & $A$, $\Phi$ & $m=\MassScanRange$~GeV (\MassScanN{} values), $\tan\beta=\MassScanTanbList$ \\
$\tan\beta$ scan & $A$, $\Phi$ & $\tan\beta=\TanbScanRange$ (\TanbScanN{} values), $m=\TanbScanMasses$~GeV \\
inclusive, Table~\ref{tab:A-inclusive} & $A$ & the $m_A=\BenchMass$~GeV points of the $\tan\beta$ scan at $\tan\beta=\InclusiveTanbList$ \\
subprocesses, Table~\ref{tab:A-subprocesses} & $A$, $h$ & $m=\BenchMass$~GeV, $\tan\beta=1$ (top) and \InclusiveTanbMax{} ($t+b$) \\
differential & $A$, $\Phi$, $h$ & $m=\BenchMass$~GeV, $\tan\beta=\DistTanbList$ \\
\hline\hline
\end{tabular}
 \end{table}

The differential samples of
Secs.~\ref{subsec:finite_mass_distributions}--\ref{subsec:phi_distributions}
combine, for each of the \DistCombinedRuns{} parameter points,
\DistRunsPerPoint{} independent integrations with different random-number seeds,
from two separate builds of the same code on two computing clusters, combined
bin by bin by inverse-variance weighting. Bin by bin the runs of each point
are statistically compatible, once allowance is made for the known
underestimate of the VEGAS errors. Their total cross sections scatter by
more than that: the mean of one batch of runs departs from the combined value
by up to $\DistBatchSpread\%$, far below every other uncertainty in this
paper but well above the quoted integration errors. The errors given for
combined totals are therefore scaled by $\sqrt{\chi^2/\text{ndf}}$ of the
runs wherever this exceeds one, by at most a factor $\DistScaleMax$. Individual runs entering the combination ran for between
\DistFullHoursLo{} and \DistFullHoursHi{} hours. The resulting integration
precision on the total cross section is \DistFullPrecision\% for the exact
top-plus-bottom samples and \DistEffPrecision\% for the \DistEffSamples{} of
EFT+FF. The points of the parameter scans, run at lower statistics with the
same inputs, have an integration precision of \ScanPrecision\%; at
$m=\BenchMass$~GeV they agree with the samples of Table~\ref{tab:dist-xsec}
to within \ScanDistDevMax\%.

Table~\ref{tab:dist-precision} gives, for each $A$ sample, that precision together
with the median per-bin uncertainty of every observable drawn in this paper.
The angular observables and $z^\star$ are the most finely resolved, with
median per-bin uncertainties at or below \DistSmoothMax\%, and the dijet
invariant mass reaches \DistMjjMedian\%. For $|\Delta\eta_{j_1j_2}|$,
$z^\star$ and $\Phi_2$, whose normalized shapes were found in
Sec.~\ref{sec:results} to be insensitive to the heavy-quark treatment at
$\tan\beta=1$, this precision shows that the agreement is not a statistical
artefact.

\begin{table}[htbp]
\centering
\caption{Integration precision of each differential $A$ sample and the median
per-bin uncertainty of each observable, for the combined samples that the
figures draw.}
\label{tab:dist-precision}
\begin{tabular}{l c c c ccccccc}
\hline\hline
sample & $\tan\beta$ & $n$ & $\delta\sigma/\sigma$ [\%]
& $\Delta\eta_{j_1j_2}$ & $m_{j_1j_2}$ & $z^\star$ & $\Phi_{2}$ & $p_{T,j_1}$ & $p_{T,j_3}$ & $p_{T,A}$
\\
& & & & \multicolumn{7}{c}{median per-bin uncertainty [\%]} \\
\hline
EFT+FF & $1$ & 12 & 0.014
& 0.12
& 0.35
& 0.25
& 0.15
& 0.20
& 0.52
& 0.40\\
full $t+b$ & $1$ & 12 & 0.030
& 0.18
& 0.53
& 0.35
& 0.21
& 0.39
& 0.87
& 0.78\\
full $t+b$ & $7$ & 12 & 0.027
& 0.18
& 0.63
& 0.37
& 0.22
& 0.51
& 1.32
& 1.15\\
full $t+b$ & $10$ & 12 & 0.022
& 0.19
& 0.65
& 0.38
& 0.24
& 0.65
& 1.81
& 1.33\\
full $t+b$ & $15$ & 12 & 0.023
& 0.19
& 0.66
& 0.40
& 0.25
& 0.85
& 2.26
& 1.59\\
full $t+b$ & $30$ & 12 & 0.030
& 0.20
& 0.69
& 0.40
& 0.26
& 0.95
& 2.69
& 1.73\\
\hline\hline
\end{tabular}
 \end{table}

The transverse-momentum tails are the exception, and they are the reason the
distributions in Secs.~\ref{subsec:finite_mass_distributions} and
\ref{subsec:thirdjet} are drawn with combined bins there. Above
$\TailPtHiggsCut$~GeV the median per-bin uncertainty of the $p_{T,A}$
spectrum is \TailPtHiggsNative\% in the native binning; combining
\RebinPtHiggsFactor{} bins brings it to \TailPtHiggsRebinned\%. Above
$\TailPtjthreeCut$~GeV the corresponding figures for $p_{T,j_3}$ are
\TailPtjthreeNative\% and \TailPtjthreeRebinned\%.
Table~\ref{tab:rebinning} states the rule applied to each observable; in
$\Phi_2$, which is populated almost uniformly, \RebinPhitwojjFactor{} bins are
combined everywhere.

\begin{table}[htbp]
\centering
\caption{Bin-combination rule applied to each observable when drawn.}
\label{tab:rebinning}
\begin{tabular}{ll}
\hline\hline
observable & bin combination \\
\hline
$\Delta\eta_{j_1j_2}$ & native binning \\
$m_{j_1j_2}$ & native binning \\
$p_{T,j_3}$ & 5 bins, $x>110$ \\
$p_{T,j_1}$ & 5 bins, $x>400$ \\
$z^\star$ & 2 bins, everywhere \\
$p_{T,A}$ & 5 bins, $x>600$ \\
$\Phi_{2}$ & 4 bins, everywhere \\
\hline\hline
\end{tabular}
 \end{table}

\clearpage
\bibliographystyle{h-physrev}
\bibliography{paper-h}

\end{document}